\documentclass[11pt]{article}
 \usepackage{amsmath, amsthm, amssymb, bbm, setspace,bigints}
 \usepackage[margin=1 in]{geometry}
\usepackage{caption}
\usepackage{subcaption}
\usepackage[toc,page]{appendix}     
\usepackage{pdfpages}
\usepackage{epstopdf}
\usepackage{booktabs}
\usepackage{tabularx, multirow}
\usepackage[square,sort,comma,numbers]{natbib}
\usepackage{empheq}
\usepackage{algorithmic}
\usepackage{algorithm}

\newcommand{\be} {\begin{eqnarray*}}
\newcommand{\ee} {\end{eqnarray*}}

\newcommand{\argmin}{\mathop{\rm argmin~}}
\newcommand{\vep}{\varepsilon}
\newcommand{\thetahat}{\widehat{\theta}}
\newcommand{\Xtil}{\widetilde{X}}
\newcommand{\gammahat}{\widehat{\gamma}}

\newcommand{\Deltahat}{\widehat{\Delta}}
\newcommand{\Thetahat}{\widehat{\Theta}}
\newcommand{\Sigmahat}{\widehat{\Sigma}}
\newcommand{\thetastar}{\theta^\ast}
\newcommand{\gammastar}{\gamma^\ast}
\newcommand{\rhobar}{\overline{\rho}}
\newcommand{\alphastar}{\alpha^\ast}

\def\m{\mathcal}
\def\mb{\mathbb}

\newtheorem{theorem}{Theorem}
\newtheorem{lemma}[theorem]{Lemma}

\newtheorem{proposition}[theorem]{Proposition}
\newtheorem{corollary}[theorem]{Corollary}

\title{Statistical inference for high dimensional regression \\ via Constrained Lasso}
\author{Yun Yang\\
Florida State University}   
\date{\vspace{-1em}}

\begin{document}
\maketitle

\begin{abstract}
In this paper, we propose a new method for estimation and constructing confidence intervals for low-dimensional components in a high-dimensional model. The proposed estimator, called Constrained Lasso (CLasso) estimator, is obtained by simultaneously solving two estimating equations---one imposing a zero-bias constraint for the low-dimensional parameter and the other forming an $\ell_1$-penalized procedure for the high-dimensional nuisance parameter. By carefully choosing the zero-bias constraint, the resulting estimator of the low dimensional parameter is shown to admit an asymptotically normal limit attaining the Cram\'{e}r-Rao lower bound in a semiparametric sense. We propose a tuning-free iterative algorithm for implementing the CLasso. We show that when the algorithm is initialized at the Lasso estimator, the de-sparsified estimator proposed in van de Geer et al. [\emph{Ann. Statist.} {\bf 42} (2014) 1166--1202]
is asymptotically equivalent to the first iterate of the algorithm. We analyse the asymptotic properties of the CLasso estimator and show the globally linear convergence of the algorithm.  We also demonstrate encouraging empirical performance of the CLasso through numerical studies.

\end{abstract}

\section{Introduction}
Various statistical procedures have been proposed over the last decade for solving high dimensional statistical problems, where the dimensionality of the parameter space may exceed or even be much larger than the sample size. Under certain low-dimensional structural assumption such as sparsity, the high-dimensional problem becomes statistically identifiable and estimation procedures are constructed in various ways to achieve estimation minimax optimality \cite{tibshirani1996regression,bickel2009simultaneous,zhang2008sparsity,
van2008high,yuan2006model,friedman2008sparse,zou2005regularization,tibshirani2005sparsity} and variable selection consistency \cite{zhao2006model,meinshausen2006high,wainwright2009sharp,yang2016computational}.
See the book~\cite{buhlmann2011statistics} and the survey article~\cite{fan2010selective} for a selective review on this subject.

On the other hand, due to the intractable limiting distribution of sparsity-inducing estimators such as the Lasso \cite{tibshirani1996regression}, little progress has been made on how to conduct inference. Vanilla bootstrap and subsampling techniques fail to work for the Lasso
even in the low-dimensional regime due to the non-continuity and the unknown parameter-dependence of the limiting distribution~\cite{knight2000asymptotics}. Moreover, Leeb has shown in a series of his work~\cite{leeb2005model,leeb2006can,leeb2008can} that there is no free lunch---one cannot achieve super-efficiency and accurate estimation of the sampling distribution of the super-efficient estimator at the same time.
Recently, initiating by the pioneer work \cite{zhang2014,van2014,javanmard2014confidence}, people start seeking point estimators in high-dimensional problems that are not super-efficient but permit statistical inference, such as constructing confidence intervals and conducting hypothesis testing. Reviews and comparisons regarding other statistical approaches for quantifying uncertainties in high-dimensional problems can be found in \cite{van2014} and \cite{dezeure2015high}.

In \cite{zhang2014,van2014,javanmard2014confidence}, they propose a class of de-biased, or de-sparsified estimators by removing a troublesome bias term due to penalization that prevents the root-$n$ consistency of the Lasso estimator. 
This new class of post-processing estimators are no longer super-efficient but shown to admit asymptotically normal limiting distributions, which facilitates statistical inference. However, as we empirically observed in the numerical experiments, this solution is still not satisfactory since confidence intervals based on these de-sparsified estimators tend to be under-coveraged  for unknown signals with non-zero true values, meaning that the actual coverage probabilities of the confidence intervals are lower than their nominal significance level; and tend to be over-coveraged for zero unknown true signals, meaning that the actually coverage probability higher than nominal. This unappealing practical performance of de-sparsified estimators can be partly explained by the somehow crude de-biasing procedure for correcting the bias, as the ``bias-corrected" estimator has not been fully escaped from super-efficiency.

In this paper, we take a different route by directly imposing a zero-bias constraint for the low-dimensional parameter $\theta$ of interest accompanied by an $\ell_1$-penalized procedure for estimating the remaining high--dimensional nuisance parameter $\gamma$. This new zero-bias constraint requires the projection of the fitted residual of the response vector onto certain carefully chosen directions to vanish.
From a semiparametric perspective, a carefully chosen constraint has the effect of forcing the efficient score function along certain least favourable submodel to be close to zero when evaluated near the truth. We show that the resulting \emph{Constrained Lasso} (CLasso) estimator admits an asymptotically normal limit and achieves optimal semiparametric efficiency, meaning that its asymptotic covariance matrix attains the semiparametric Cram\'{e}r-Rao lower bound. We propose an iterative algorithm for numerically computing the CLasso estimator via iteratively updating $\theta$ via solving a linear system, and updating  $\gamma$ via solving a Lasso programming. The algorithm enjoys globally linear convergence up to the statistical precision of the problem, meaning the typical distance between the sampling distribution of the estimator and its asymptotic normal distribution.   Different from gradient-based procedures where the optimization error typically contracts at a constant factor independent of sample size $n$ and dimensionality $p$ (but depends on the conditional number) of the problem, our algorithm exhibits a contraction factor proportional to $\sqrt{(s^2/n)\log p}$ (this quantity encodes the typical difficulty of high-dimensional statistical problems with sparsity level $s$) that decays towards zero as $(s^2/n)\log p \to 0$. Moreover, our algorithm involves no step size and is tuning free. In our numerical experiments, a few iterations such as $10$ are typically suffice for the algorithm to well converge.

More interestingly, we find a close connection between the CLasso and the aforementioned de-sparsified procedures proposed in \cite{van2014}---when initialized at the Lasso estimator, the de-sparsified estimator is shown to be asymptotically equivalent to the first iterate from our iterative algorithm for solving the CLasso. This close connection explains and solves the under- and over-coverage issue associated with the de-sparsified estimator: by refining the de-sparsified estimator through more iterations, the resulting CLasso estimator is capable of escaping from the super-efficiency region centered around the Lasso initialization, and leads to more balanced coverages for truth unknown signals with both zero and non-zero values. Depending on the convergence speed of the algorithm, the improvement on the coverage can be significant---this also suggests a poor performance of the de-sparsified estimator when algorithmic rate of convergence $\sqrt{(s^2/n)\log p}$ is large.

Overall, our results suggest that by incorporating constraints with widely-used high dimensional penalized methods, we are able to remove the bias term appearing in limiting distributions of low-dimensional components in high-dimensional models that prevents us from conducting statistical inference at the price of losing super-efficiency. By carefully selecting the constraints, we can achieve the best efficiency in the semiparametric sense. 

The rest of the paper is organized as follows. In Section~\ref{Sec:Main}, we motivate and formally introduce the CLasso method. We also propose an iterative algorithm for implementing the CLasso, and discuss its relation with de-sparsified estimators. In Section~\ref{Sec:Theory}, we provide theory of the proposed method, along with a careful convergent analysis of the iterative algorithm. In Section~\ref{Sec:simulation}, we conduct numerical experiments and apply our method to a real data. We postpone all the proofs to Section~\ref{Sec:Proof} and conclude the paper with a discussion in Section~\ref{Sec:Discussion}.

\section{Constrained Lasso}\label{Sec:Main}
To begin with, we formulate the problem and describe the key observation in Section~\ref{Sec:Motivation} that motivates our method. In Section~\ref{Sec:methods}, we formally introduce the new method, termed Constrained Lasso (CLasso), proposed in this paper. In Section~\ref{Sec:Algorithm}, we describe an iterative algorithm for implementing the CLasso. In Section~\ref{Sec:Relation}, we illustrate a close connection between the proposed method and a class of de-sparsifying based methods.

\subsection{Motivation}\label{Sec:Motivation}
Consider the linear model:
\begin{align}\label{Eqn:LM}
Y = U \beta + w, \qquad  w\sim \m N(0,\,\sigma^2I_n),
\end{align}
where $U=(X,\, Z)\in\mb R^{n\times (d+p)}$ is the design matrix, $\beta = (\theta^T,\,\gamma^T)^T\in\mb R^{d+p}$ is the unknown regression coefficient vector, $Y\in\mb R^n$ is the response vector and $w$ is a Gaussian noise vector. Suppose among all components of $\beta$, we are only interested in conducting statistical inference for its first $d$ components, denoted by $\theta\in\mb R^d$, and the remaining part $\gamma\in\mb R^p$ is a nuisance parameter. Correspondingly, we divide the design matrix $U$ into two parts: design matrix $X\in\mb R^{n\times d}$ for the parameter of interest and design matrix $Z\in\mb R^{n\times p}$ for the nuisance part. Under this setup, we can rewrite the model into a semiparametric form:
\begin{align}\label{Eqn:Linear_Model}
Y = X\theta + Z\gamma +w,\qquad w\sim\m N(0,\, \sigma^2I_n).
\end{align}

We are interested in the regime where the nuisance parameter is high-dimensional, or $p\gg n$, while the parameter of interest is low-dimensional, or $d\ll n$. A widely-used method for estimating the regression coefficient $\beta$, or the $(\theta,\,\gamma)$ pair, is the Lasso~\cite{tibshirani1996regression},
\begin{align}\label{Eqn:LASSO}
(\thetahat_L,\,\gammahat_L)=\argmin_{\theta\in\mb R^d,\,\gamma\in\mb R^p} \Big\{\frac{1}{2n}\,\|Y-X\theta-Z\gamma\|^2 + \lambda\, \|\theta\|_1+\lambda\, \|\gamma\|_1\Big\},
\end{align}
where $\lambda$ is a regularization parameter controlling the magnitude of the $\ell_1$-penalty term in the objective function. Under the assumption that the true unknown regression coefficient vector $\beta^\ast$ is $s$-sparse with $s\ll n$, the optimal scaling of regularization parameter $\lambda$ is of order $\sqrt{n^{-1}\log p}$, leading to minimax-rate $\sqrt{(s/n)\log p}$ of estimation and prediction \cite{raskutti2011minimax}. However, due to the $\ell_1$-penalty term, the resulting estimator $\thetahat_L$ is biased, with a bias magnitude proportional to $\lambda$ (see, for example, \cite{knight2000asymptotics} for fix-dimensional results and \cite{wainwright2009sharp} for high-dimensional results). This $\sqrt{n^{-1}\log p}$-magnitude bias destroys the root $n$-consistency of $\theta$ as the dimensionality $p$ grows with $n$, rendering statistical inference for $\thetahat_L$ an extremely difficult task.
The most relevant method in the literature is a class of post-processing procedures developed in \cite{zhang2014,van2014,javanmard2014confidence}, where an estimator of $\theta$ is constructed by removing from the original Lasso estimator $\thetahat_L$ an``estimated bias term" that prevents $\thetahat_L$ from achieving the root-$n$ consistency. As we discussed in the introduction, this post-processing procedure tends to have unappealing empirical performance due to the seemingly crude bias-correction when the original statistical problem is hard, meaning that $\sqrt{(s^2/n)\log p}$ is relatively large.

In this work, we take a different route by directly removing the bias term through combining a bias-eliminating constraint with the Lasso procedure~\eqref{Eqn:LASSO}. This new approach deals with the bias directly and is free of post-processing. Surprisingly, as we will show in Section~\ref{Sec:Relation}, the de-sparsified Lasso estimator proposed in \cite{van2014} is asymptotically equivalent to the first iterate in our algorithm (Algorithm.~\ref{Alg:BFLasso}) for solving the Constrained Lasso (CLASSO). 

To motivate the method, let us first consider a naive correction to the original Lasso programming~\eqref{Eqn:LASSO} by removing the penalty term of $\theta$, which will be referred to as the un-penalized Lasso (UP Lasso),
\begin{align}\label{Eqn:UPLASSO}
(\thetahat_U,\,\gammahat_U)=\argmin_{\theta\in\mb R^d,\,\gamma\in\mb R^p} \Big\{\frac{1}{2n}\,\|Y-X\theta-Z\gamma\|^2 +\lambda\, \|\gamma\|_1\Big\}.
\end{align}
It can be shown that the KKT condition of the UP Lasso is
\begin{subequations}\label{Eqn:Z_est_UP}
\begin{empheq}[left=\empheqlbrace]{align} 
&\frac{1}{n}\, X^T(Y - X \theta-Z\gamma) = 0 \label{Eqn:ZUPa}\\
&\frac{1}{n}\, Z^T (Y - X \theta - Z \gamma) = \lambda \,\kappa,\quad \kappa\in \partial\|\gamma\|_1.\label{Eqn:ZUPb}
\end{empheq}
\end{subequations}
By plugging in $Y=X\thetastar + Z\gammastar+w$, where $(\thetastar,\,\gammastar)$ denotes the true parameter, and rearranging the terms, the first KKT condition on $\theta$ can be rearranged as
\begin{align}\label{Eqn:theta_UP}
\sqrt{n}\,(\thetahat_U-\thetastar) =  \frac{1}{\sqrt{n}}\, (X^TX)^{-1} \,X^Tw + \frac{1}{\sqrt{n}}\, (X^TX)^{-1} \, X^TZ\,(\gammahat_U-\gammastar).
\end{align}
Under some reasonable assumption on $X$, the first term converges to a normal limit, while the second term has a typical order $\sqrt{s\,\log p}$ that does not vanish as $n$ increases. As a consequence, the UP Lasso estimate $\thetahat_U$ still fails to achieve the root $n$-consistency of $\theta$.

After taking a more careful look at the decomposition~\eqref{Eqn:theta_UP}, we find that the second bias term will be exactly zero if columns of $X$ are orthogonal to columns of $Z$ (or designs of $X$ and $Z$ are orthogonal). This suggests that the last bias term in $\thetahat_U$ is primarily due to the non-zero projection of the bias $Z(\gammahat_U-\gammastar)$ in the nuisance part onto the column space of $X$.
Consequently, if we replace the KKT condition~\eqref{Eqn:ZUPa} on $\theta$  by
\begin{align*}
\frac{1}{n}\, (X-Z\alpha)^T(Y - X \theta-Z\gamma) = 0,
\end{align*}
for some suitable matrix $\alpha\in\mb R^{p\times d}$ such that product $(X-Z\alpha)^T Z$ is close to zero in some proper metric, then the same argument leads to
\begin{align*}
\sqrt{n}\,(\thetahat-\thetastar) =  \frac{1}{\sqrt{n}}\, (\Xtil^TX)^{-1} \,\Xtil^Tw + \frac{1}{\sqrt{n}}\, (\Xtil^TX)^{-1} \, \Xtil^TZ\,(\gammahat-\gammastar),
\end{align*}
where we use $\Xtil = X-Z\alpha$ to denote the residual of $X$ after subtracting $Z\alpha$. Since now $(X-Z\alpha)^T Z$ is close to zero, the second bias term will be vanished as $n\to\infty$, leading to the asymptotic normality of $\thetahat$. This observation motivates the new method proposed in the next subsection.

\subsection{Methods}\label{Sec:methods}

According to the observations in the previous subsection, we propose a new high-dimensional $Z$-estimator of $(\thetahat,\,\gammahat)$ that simultaneous solves the following two estimation equations that are obtained via modifying the KKT conditions~\eqref{Eqn:Z_est_UP} of the UP Lasso,
\begin{subequations}\label{Eqn:Z_est}
\begin{empheq}[left=\empheqlbrace]{align} 
&\frac{1}{n}\, (X-Z\alpha)^T(Y - X \theta-Z\gamma) = 0 \label{Eqn:Za}\\
&\frac{1}{n}\, Z^T (Y - X \theta - Z \gamma) = \lambda \,\kappa,\quad \kappa\in \partial\|\gamma\|_1,\label{Eqn:Zb}
\end{empheq}
\end{subequations}
where the construction of the critical matrix $\alpha\in\mb R^{p\times d}$ will be specified later. At this moment, we can simply view $\alpha$ as some ``good" matrix that makes the product $(X-Z\alpha)^TZ$ close to a zero matrix.
The second equation~\eqref{Eqn:Zb} corresponds to the KKT condition of the Lasso programming of regressing the residual $Y-X\theta$ on $Z$. Therefore, problem~\eqref{Eqn:Z_est} can be expressed into an equivalent form as
\begin{subequations}\label{Eqn:M_est}
\begin{empheq}[left=\empheqlbrace]{align} 
&\frac{1}{n}\, (X-Z\alpha)^T(Y - X \theta-Z\gamma) = 0 \label{Eqn:Ma}\\
&\gamma \in \argmin_{r\in\mb R^p} \Big\{\frac{1}{2n}\,\|Y-X\theta-Zr\|^2 + \lambda\,\|r\|_1\Big\}.\label{Eqn:Mb}
\end{empheq}
\end{subequations}
We say columns of $Z$ is in a general position, or simply $Z$ is in a general position, if the affine span of any $k+1\leq n$ points $\{s_1Z_{j_1},\ldots,s_kZ_{j_k}\}$, for arbitrary signs $s_1,\ldots, s_k\in \{-1, 1\}$, does not contain any element of $\{\pm X_j: \, j\neq j_1,\ldots,j_k\}$. In many examples, such as when entries of $Z\in\mb R^{n\times p}$ is drawn from a continuous distribution on $\mb R^{n\times p}$, $Z$ satisfies this condition.
When $Z$ is in a general position, for any $\theta$ the residual regression equation~\eqref{Eqn:Mb} has a unique solution, which is also the unique solution of equation~\eqref{Eqn:Zb} \cite{tibshirani2013lasso}. Therefore, both equations~\eqref{Eqn:Z_est} and \eqref{Eqn:M_est} are well-posted. 

We will refer to these two equivalent methods of estimating $\theta$ as \emph{Constrained Lasso} (CLasso). In the special case of $\alpha=0$, equation~\eqref{Eqn:Z_est} coincides with the KKT condition of the UP Lasso problem~\eqref{Eqn:UPLASSO} and therefore UP Lasso~\eqref{Eqn:UPLASSO} is a special case of the CLasso with $\alpha\equiv 0$.  
However, for general $\alpha$ matrices, equation~\eqref{Eqn:Z_est} may not correspond to the KKT condition of any optimization problem.
The following theorem show that when $Z$ is in a general position, the CLasso is a well-posed procedure with a unique solution with a high probability.

\begin{theorem}\label{Thm:Uniqueness}
Suppose the assumptions in Section~\ref{Sec:Normality} holds. In addition suppose we have $\mu\geq \widetilde{C}\,\tau\,s$ for some constant $\widetilde{C}$ independent of $(n,\,p,\,s)$ (for precise definitions of those quantities, please refer to Section~\ref{Sec:Normality}), and $Z$ is in a general position and satisfies the sparse eigenvalue condition (SEC): $n^{-1}\,\|Z\,u\|^2 \geq \mu\,\|u\|^2$ for all vectors $u$ with sparsity level $C's$ for some sufficiently large constant $C'$. Then under the same choice of $\lambda$ as in Theorem~\ref{Thm:Normality}, we have that with probability at least $1-p^{-c}-n^{-c}$ for some $c>0$, the estimating equation~\eqref{Eqn:Z_est} or equation~\eqref{Eqn:M_est} admits a unique solution.
\end{theorem}

\noindent According to results in Section~\ref{Sec:Normality}, $(\mu,\,C)$ are constants and $\tau$ is typically of order $\sqrt{n^{-1}\log p}$. Consequently, the additional assumption $\mu\geq \widetilde{C}\,\tau\,s$ is always satisfied as long as we are in the regime $s\sqrt{n^{-1}\log p} \ll 1$, where recall that $s$ is the sparsity of the true unknown nuisance parameter $\gammastar$. This assumption for statistical inference in high dimensional sparse linear regression turns out to be stronger than common sufficient condition $\sqrt{(s/n)\log p} \ll 1$ for estimation consistency (see \cite{cai2015confidence,javanmard2015biasing} for some detailed discussions concerning these conditions). The sparse eigenvalue condition is also stronger than the restricted eigenvalue condition made in Section~\ref{Sec:Normality} for proving Lasso consistency, although both of them can be verified for a class of random design matrices \cite{raskutti2010restricted}.

Now we specify our choice for the critical matrix $\alpha$. 
Let us introduce some notation first. For any $m$ by $n$ matrix $A=(A_{ij})_{m\times n}$, we use $A^i$ to denote the $i$th row of $A$ and $A_j$ its $j$th column.  
For any vector $a\in\mb R^m$ and any index set $T\subset \{1,2,\ldots,p\}$, we use the shorthand $a_{T}$ to denote the vector formed by keeping the components whose indices are in $T$ unchanged and setting the rest to be zero. 
From now on, we always assume the design to be random with zero mean, meaning that rows $\{X^i\}_{i=1}^n$ and $\{Z^i\}_{i=1}^n$ of the two design matrices $X$ and $Z$ are i.i.d.~random vectors with dimensions $d$ and $p$, respectively, and satisfy $\mb E[X^i]=0$ and $\mb E[Z^i]=0$. Denote the covariance matrix of $U^i=(X^i,\,Z^i)$ by
\begin{align*}
\Sigma=\mb E\big[ (U^i)^T U^i\big] =  \left[\begin{array}{cc} E\big[ (X^i)^T X^i\big] & E\big[ (X^i)^T Z^i\big]\\ E\big[ (Z^i)^T X^i\big] & E\big[ (Z^i)^T Z^i\big]\\ \end{array}\right] =\left[\begin{array}{cc} \Sigma_{X,X} & \Sigma_{X,Z}\\ \Sigma_{Z,X} & \Sigma_{Z,Z}\\ \end{array}\right].
\end{align*}
Under the random design assumption, the ideal choice of $\alpha\in\mb R^{p\times d}$ would be 
\begin{align}\label{Eqn:OPTalpha}
\alphastar=\Sigma_{Z,Z}^{-1}\,\Sigma_{Z,X}=\big\{\mb E [(Z^i)^TZ^i]\big\}^{-1}\mb E[(Z^i)^TX^i],
\end{align}
since it satisfies the population level uncorrelated condition $\mb E\big[(X-Z\alphastar)^TZ\big]=\sum_{i=1}^n \mb E\big[(X^i-Z^i\alphastar)^TZ^i\big]=0$, from which we may expect its empirical version $n^{-1}(X-Z\alphastar)^TZ=n^{-1}\,\sum_{i=1}^n (X^i-Z^i\alphastar)^TZ^i$ to be close to zero with a high probability. To motivate our constructing procedure for $\alpha$, we use another equivalent definition of $\alphastar$ as the minimizer of $\mb E\big[\|X^i - Z^i\alpha\|^2\big]$, or equivalently, for each $j\in\{1,2,\ldots,d\}$, the $j$th column $\alphastar_j$ is the minimizer of the mean squared residual $\mb E\big[|X_{ij} - Z^i\alpha_j|^2\big]$, where recall that $X_{ij}=X^i_j$ denotes the $(i,j)$th component of any matrix $X$. In practice, these population level quantities are rarely known and we propose to estimate each column $\alpha_j$ via the following node-wise regression \cite{meinshausen2006high} by minimizing a penalized averaging squared residuals,
\begin{align}\label{Eqn:Choose_alpha}
\alpha_j =\argmin_{a\in\mb R^p} \, \Big\{\frac{1}{2n}\,\|X_j-Za\|^2+\lambda_j\,\|a\|_1\Big\}.
\end{align}

The CLasso has an interpretation from the semiparametric efficiency theory.  Let $\mb P_{\theta,\,\gamma}$ denote the probability distribution of the linear model~\eqref{Eqn:Linear_Model} with parameter pair $\{\theta,\,\gamma\}$. When $X$ is not orthogonal to $Z$, $0$ is not the least favourable direction (see the following for a brief explanation) of the nuisance part for estimating $\theta$. Therefore, equation~\eqref{Eqn:Za} is not the right constraint to impose.
In fact, in model~\eqref{Eqn:Linear_Model}, the (multivariate) least favourable direction is given by $\alphastar=\Sigma_{Z,Z}^{-1}\,\Sigma_{Z,X}$ (the same $\alphastar$ as previously defined in~\eqref{Eqn:OPTalpha}), meaning that $d$-dimensional the sub-model $\m P_S=\{\mb P_{\theta_t,\,\gamma_t}:\, \theta_t=t,\,\gamma_t=\gammahat + \alphastar(t-\thetahat),\, t\in\mb R^d\}$ is the hardest parametric sub-problem within the original statistical distribution family $\{\mb P_{\theta,\,\gamma}:\, \theta\in\mb R^d,\,\gamma\in\mb R^p\}$ that passes through $\mb P_{\thetahat,\,\gammahat}$. 
This parametric sub-model achieves the semiparametric Cram\'{e}r Rao lower bound of the asymptotic variance of any asymptotically unbiased estimator of $\theta$, which is $\big\{\mb E\big[(X^i-Z^i\alphastar)^T(X^i-Z^i\alphastar)\big]\big\}^{-1}=\big(\Sigma_{X,X}-\Sigma_{X,Z}\Sigma_{Z,Z}^{-1}\Sigma_{Z,X}\big)^{-1}$ (this can be proved by applying the Gauss-Markov theorem, see Section 2.3.3 in \cite{van2014} for a rigorous statement and more details). Therefore, in order to achieve the best asymptotic efficiency, we need focus on the score function (derivative of the negative likelihood function) along the path in the least favourable sub-model $\m P_S$ (the corresponding score function is called efficient score function), 
\begin{align*}
\frac{\partial}{\partial t} \Big\{\frac{1}{2n}\,\|Y-X\theta_t-Z\gamma_t\|^2\Big\} = \frac{1}{n}\,(X-Z\alphastar)^T(Y-X\theta_t-Z\gamma_t),
\end{align*}
and enforce it to be zero at $t=0$ to remove the bias, that is, by requiring
\begin{align*}
\frac{1}{n}\,(X-Z\alphastar)^T(Y-X\thetahat-Z\gammahat)=0.
\end{align*}
This constraint can be interpreted as force the impact of the bias $Z(\gammahat-\gammastar)$ from the nuisance part on the least favourable sub-model to vanish. When $\alphastar$ is not directly available, we may again replace it with any approximation $\alpha$  under which $(X-Z\alpha)^TZ$, and therefore the efficient score function at $t=0$, is close to zero. This also leads to the node-wise regression procedure~\eqref{Eqn:Choose_alpha} of choosing $\alpha$ in the CLasso, where the KKT condition of the node-wise regression implies an element-wise sup-norm bound on the product $n^{-1}(X-Z\alphastar)^TZ$ (see Theorem~\ref{Thm:alpha}).
This second semiparametric interpretation heuristically explains the optimality of the CLasso in terms of achieving the smallest asymptotic variance (for a rigorous statement, see Section 2.3.3 in \cite{van2014} and Corollary~\ref{Cor:SemiOpt}).

\subsection{Iterative algorithm for solving the Constrained Lasso}\label{Sec:Algorithm}
We propose an iterative algorithm for solving the CLasso problem~\eqref{Eqn:Z_est} and its equivalent form~\eqref{Eqn:M_est}. More specifically, we iteratively solve $\theta$ from  equation~\eqref{Eqn:Za} and $\gamma$ from equation~\eqref{Eqn:Zb} in an alternating manner. More precisely, at iteration $t$ with a current iterate $\gamma^t$ for $\gamma$, the first equation~\eqref{Eqn:Za} yields an updating formula for $\theta$ as
\begin{align*}
\theta^{t+1} =\big[ (X-Z\alpha)^TX\big]^{-1} (X-Z\alpha)^T(Y-Z\gamma^t).
\end{align*}
In the case $\alpha=0$, this reduces to $\theta^{t+1} = (X^TX)^TX^T(Y-Z\gamma^t)$, which is the least square estimate for fitting the residual $Y-Z\gamma^t$ obtained by subtracting the nuisance part from the response. Next, given the newly updated iterate $\theta^{t+1}$ for $\theta$, we update $\gamma$ by using the equivalence between equation~\eqref{Eqn:Zb} and equation~\eqref{Eqn:Mb} via
\begin{align*}
\gamma^{t+1}=\argmin_{\gamma\in\mb R^p}\Big\{\frac{1}{2n}\,\|Y - X\theta^{t+1}-Z\gamma\|^2+\lambda\,\|\gamma\|_1\Big\}.
\end{align*}
Since this optimization problem shares the same structure as the Lasso programming by equating the response variable with the current residual $Y-X\theta^{t+1}$, we can use the state-of-the-art algorithm (such as the \texttt{glmnet} package in R) of the Lasso programming to efficiently find $\gamma^{(t+1)}$. In practice, the following unadjusted Lasso estimate serves as a good initialization of the algorithm,
\begin{align*}
(\theta^0,\,\gamma^0)=\argmin_{\theta\in\mb R^d,\,\gamma\in\mb R^p} \Big\{\frac{1}{2n}\,\|Y-X\theta-Z\gamma\|^2 + \lambda\, \|\theta\|_1+\lambda\, \|\gamma\|_1\Big\}.
\end{align*}
We may also consider a more general form of the algorithm by allowing the regularization parameter $\lambda=\lambda_t$ to change across the iterations. The reason for using a $t$-dependent $\lambda$ is as follows. In order for the algorithm to have globally exponential convergence from any initialization, we need to pick a slightly larger $\lambda_t$ that grows proportionally to $\|\gamma^t-\gammastar\|_1$ at the beginning. However, a larger $\lambda_t$ tends to incur a large bias in $\gamma^t$, which in turn induces a large bias in $\theta^t$. Therefore, as $\gamma^t$ becomes close to $\gammastar$ as the algorithm proceeds, we may gradually reduce $\lambda^t$ to make it close to the optimal $\lambda$ with order $\sigma\sqrt{n^{-1}\log p}$. Rigorous analysis of the convergence of this algorithm and the associated estimation error bounds can be found in Section~\ref{Sec:Alg_rate}.
At the end of this subsection, we summarize the full algorithm for implementing the CLasso in Algorithm.~\ref{Alg:BFLasso} below.

\begin{algorithm}[!h]
\caption{CLasso Algorithm}
\label{Alg:BFLasso}
\begin{tabbing}
\enspace Input: response $Y\in\mb R^d$, design matrices $X\in\mb R^{n\times d}$ and $Z\in\mb R^{n\times p}$\\
\enspace Output: Fitted $\thetahat$ and $\gammahat$, and the asymptotic covariance matrix $\widehat{\Omega}$ of $\sqrt{n}\,(\thetahat-\thetastar)$
\enspace\\
\enspace {\bf Find matrix $\alpha$:}\\
\enspace For $j=1$ to $d$ \\
\qquad Set $\displaystyle \alpha_j =\argmin_{a\in\mb R^p} \, \Big\{\frac{1}{2n}\,\|X_j-Za\|^2+\lambda_j\,\|a\|_1\Big\}$\\
 \enspace Set $p\times d$ matrix $\alpha=(\alpha_1,\ldots,\alpha_d)$ \\
 \enspace Estimate noise variance $\widehat{\sigma}^2$ via the scaled Lasso \cite{sun2012scaled} \\
 \enspace Output $\widehat{\Omega} = \widehat{\sigma}^2\,\big[n^{-1}\,(X-Z\alpha)^T(X-Z\alpha)\big]^{-1}$\\
 \enspace \\
 \enspace {\bf Iterative algorithm for solving $\theta$:}\\
\enspace Initialize $\theta^0$ and $\gamma^0$ at the Lasso solution, that is, set\\
\enspace  $\qquad\displaystyle (\theta^0,\,\gamma^0)=\argmin_{\theta\in\mb R^d,\,\gamma\in\mb R^p}\Big\{\frac{1}{2n}\,\|Y - X\theta-Z\gamma\|^2+\lambda\,\|\theta\|_1+\lambda\,\|\gamma\|_1\Big\}$\\
 \enspace For $t=1$ to $T$  ($T$ is the number of iterations)\\
 \qquad Set $\theta^{t} =\big[ (X-Z\alpha)^TX\big]^{-1} (X-Z\alpha)^T(Y-Z\gamma^{t-1})$\\
 \qquad Set $\displaystyle \gamma^{t}=\argmin_{\gamma\in\mb R^p}\Big\{\frac{1}{2n}\,\|Y - X\theta^{t}-Z\gamma\|^2+\lambda_t\,\|\gamma\|_1\Big\}$\\
 \qquad \qquad(Here $\lambda_t\to\lambda$ as $t\to \infty$, for example, $\lambda_t=\lambda \, [1+c\,\|\gamma^{t-1}-\gamma^{t-2}\|_1]$ for $t\geq 2$,\\
 \qquad \qquad\qquad \qquad\qquad \qquad \qquad \qquad \qquad \ \ and $\lambda_1=\lambda \, [1+c\,\|\gamma^0\|_1]$)  \\
 \enspace Output solution $\thetahat=\theta^T$ and $\gammahat=\gamma^T$.
\end{tabbing}
\end{algorithm}

\subsection{Relation with De-sparsified Lasso estimator}\label{Sec:Relation}
In this subsection, we discuss the relationship between the de-sparsified Lasso estimator \cite{van2014,zhang2014} and the proposed CLasso method.
For simplicity, we consider the special case when the parameter of interest $\theta$ is one dimensional. Recall that $U=(X,\,Z)$ and $\beta = (\theta,\,\gamma)^T$ are the full design matrix and regression coefficient vector, respectively. Throughout this subsection, we consider $X\in\mb R^{n\times d}$ with $d=1$.

First, we briefly review the de-sparsified Lasso procedure. 
Following the presentation of \cite{van2014}, we denote by $\Thetahat_{L}$ an proxy of the inverse of the sample covariance matrix $\Sigmahat:\,=n^{-1}U^TU$, in the sense of making the product $\Thetahat_{L} \Sigmahat$ close to the $(d+p)$-dimensional identify matrix $I_{d+p}$. 
Their de-sparsified Lasso estimator is defined as
\begin{align*}
\widehat{b} = \widehat{\beta} + \Thetahat \, U^T\, (Y-U\,\widehat{\beta})/n,
\end{align*}
where $\widehat{\beta}$ is the unadjusted Lasso estimate, which is also our initialization $(\theta^0,\,\gamma^0)^T$ in Algorithm.~\ref{Alg:BFLasso}. Focusing on the first component of $\widehat{b}$, the parameter of interest, we express it as
\begin{align}\label{Eqn:b_1}
\widehat{b}_1 = \theta^0 + \Thetahat^1 \, \left[\begin{array}{c} X^T\\ Z^T
\end{array}\right]
\, (Y-X\theta^0-Z\gamma^0)/n,
\end{align}
where $\Thetahat^1$ denotes the first row of $\Thetahat$.
 According to \cite{van2014}, the first row $\Thetahat^1$ takes the form as
 \begin{align*}
 \Thetahat^1 = \Big(\widehat{\tau}_1^{-2},\, \widehat{\tau}_1^{-2}\,\alpha_1^T\Big),
 \end{align*}
where $\alpha_1$ is constructed in the node-wise regression~\eqref{Eqn:Choose_alpha} with $j=1$, and 
\begin{align*}
\widehat{\tau}_1 = \frac{1}{n}\,\|X-Z\alpha_1\|^2+2\, \lambda_1\,\|\alpha_1\|_1.
\end{align*}
By plugging in these into formula~\eqref{Eqn:b_1}, we obtain
\begin{align}\label{Eqn:DSLasso}
\widehat{b}_1 =  \theta^0_1 + \widehat{\tau}_1^{-2}\, (X - Z\alpha_1)^T
\, (Y-X\theta^0-Z\gamma^0)/n.
\end{align}
In comparison, it is easy to write out the updating formula for $\theta^1_1$ in the first iteration of Algorithm.~\ref{Alg:BFLasso} in a similar form
\begin{equation}\label{Eqn:BFLasso}
\begin{aligned}
&\theta^1_1 = \theta^0_1 + \widetilde{\tau}_1^{-2}\, (X - Z\alpha_1)^T
\, (Y-X\theta^0-Z\gamma^0)/n, \\
\mbox{with}\quad & \widetilde{\tau}_1^2 = \frac{1}{n}\,(X-Z\alpha_1)^T\, X.
\end{aligned}
\end{equation}
By comparing formulas~\eqref{Eqn:DSLasso} and \eqref{Eqn:BFLasso}, the only difference is in the denominator $\widehat{\tau}_1^2$ and $\widetilde{\tau}_1^2$, which are both expected to converge to the population level squared residual $\mb E\big[\|X^i-Z^i\alphastar\|^2\big]$, since $\lambda_1 \sim \sqrt{n^{-1}\log p}$ tends to be small, while both the empirical product $n^{-1}( X-Z\alpha_1)^T\, X$ and the averaging squared norm $n^{-1}\,\|X-Z\alpha_1\|^2$ tend to converge to the population level quantity $\mb E\big[(X^i-Z^i\alphastar)^TX\big] = \mb E\big[\|X^i-Z^i\alphastar\|^2\big]$ as $n\to\infty$ and $n^{-1}\log p\to 0$. More precisely, we have the following proposition.

\begin{proposition}\label{Prop:Relation}
Under the assumption in Theorem~\ref{Thm:Normality}, we have that as $n\to\infty$ and $\sqrt{s/n}\,\log p \to 0$, 
\begin{align*}
\big|\theta^1_1 - \widehat{b}_1\big| =O_P\Big(\frac{\sqrt{s} \,\log p}{n}\Big)=o_P\Big(\big|\widehat{b}_1 - \thetastar_1\big|\Big).
\end{align*}
\end{proposition}

\begin{figure}[htp]
\centering
\begin{tabular}{cc}
    \includegraphics[width=2.7in]{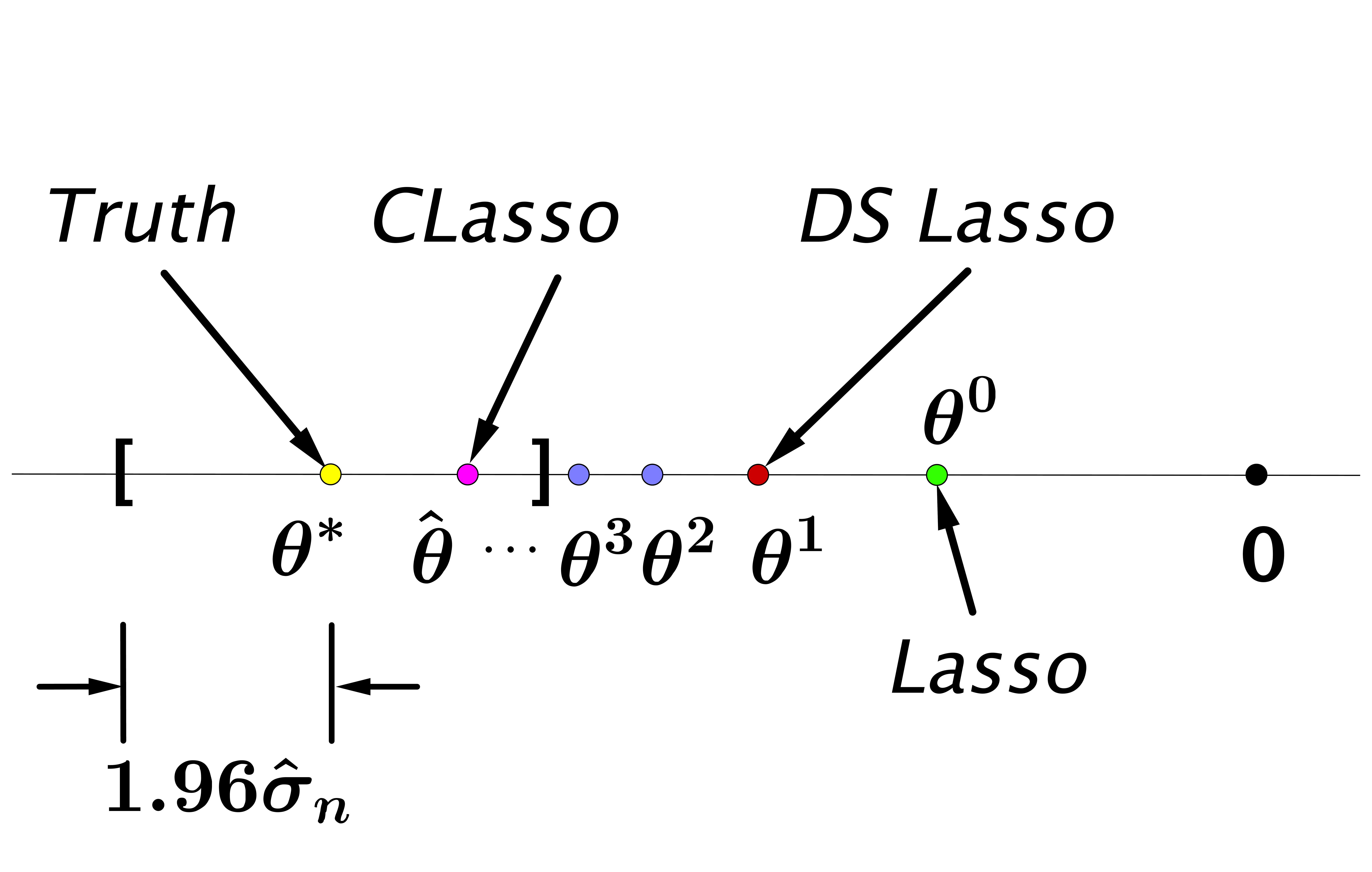}
    &
    \includegraphics[width=2.7in]{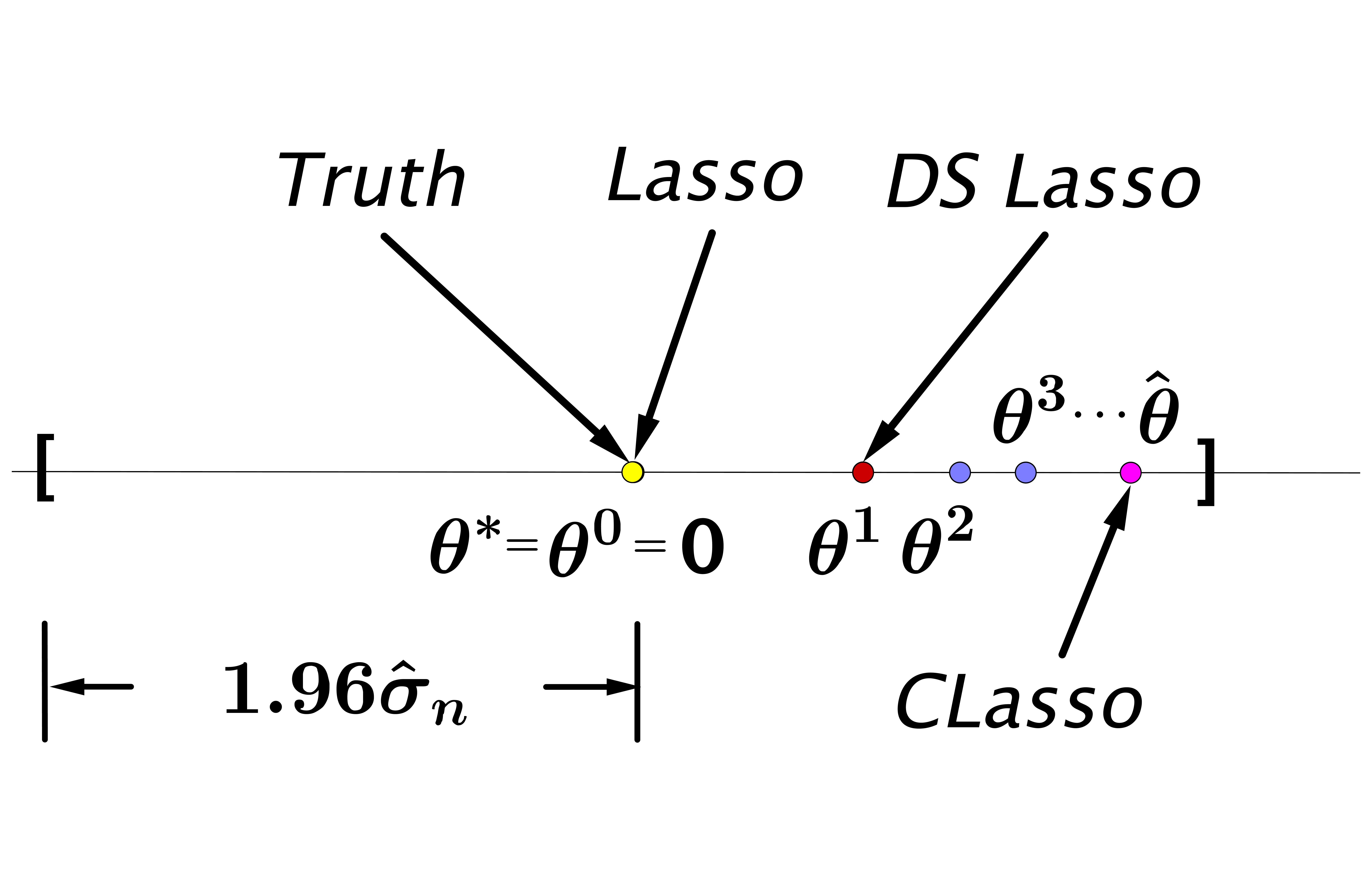}
\end{tabular}
\caption{An illustration for the relationship between the de-sparsified Lasso (DS Lasso) and the Constrained Lasso (CLasso). The left panel shows their relationship when the true signal $\theta^\ast$ is nonzero, where the Lasso estimate $\theta^0$ tends to shrink towards zero due to the $\ell_1$ penalty; and the right panel shows the relationship when $\theta^\ast$ is zero, and the Lasso estimate $\theta^0$ is also zero. The DS Lasso estimate tends to be close the to unadjusted Lasso estimate, and incurs a bias towards zero. This zero-pointing bias leads to under-coverage for a non-zero signal and over-coverage for a zero signal when confidence intervals are constructed. The square brackets indicate an interval with the same length as the $95\%$ confidence interval but centered at the truth $\thetastar$. This means, for example, in the left figure, the truth $\thetastar$ is contained in the $95\%$ confidence interval centered at the CLasso estimate $\thetahat$, but not in the confidence interval centered at the DS Lasso estimate $\theta^1$. }
\label{fig:1}
\end{figure}

\begin{figure}[htp]
\centering
\begin{tabular}{cc}
    \includegraphics[width=3in]{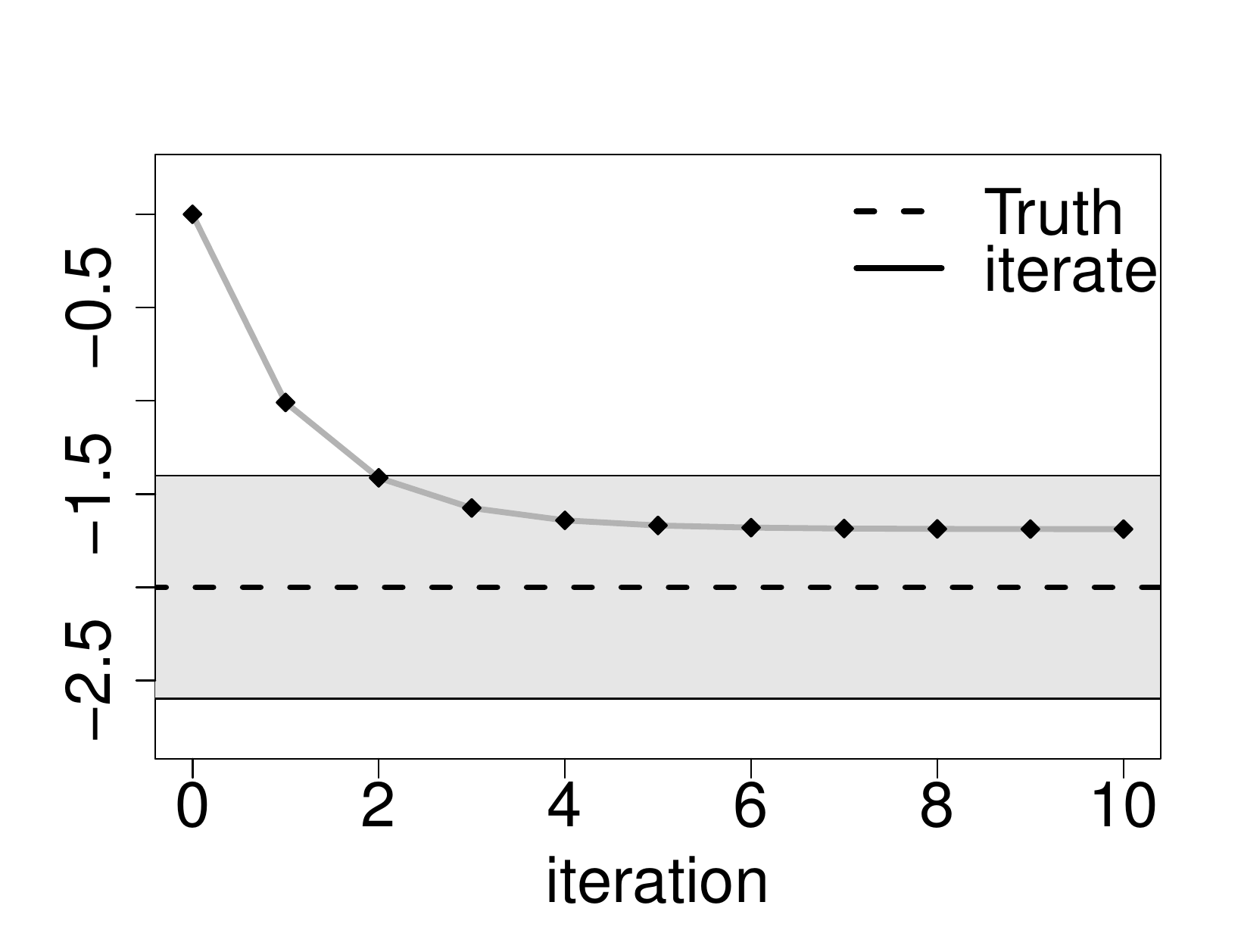}
    &
    \includegraphics[width=3in]{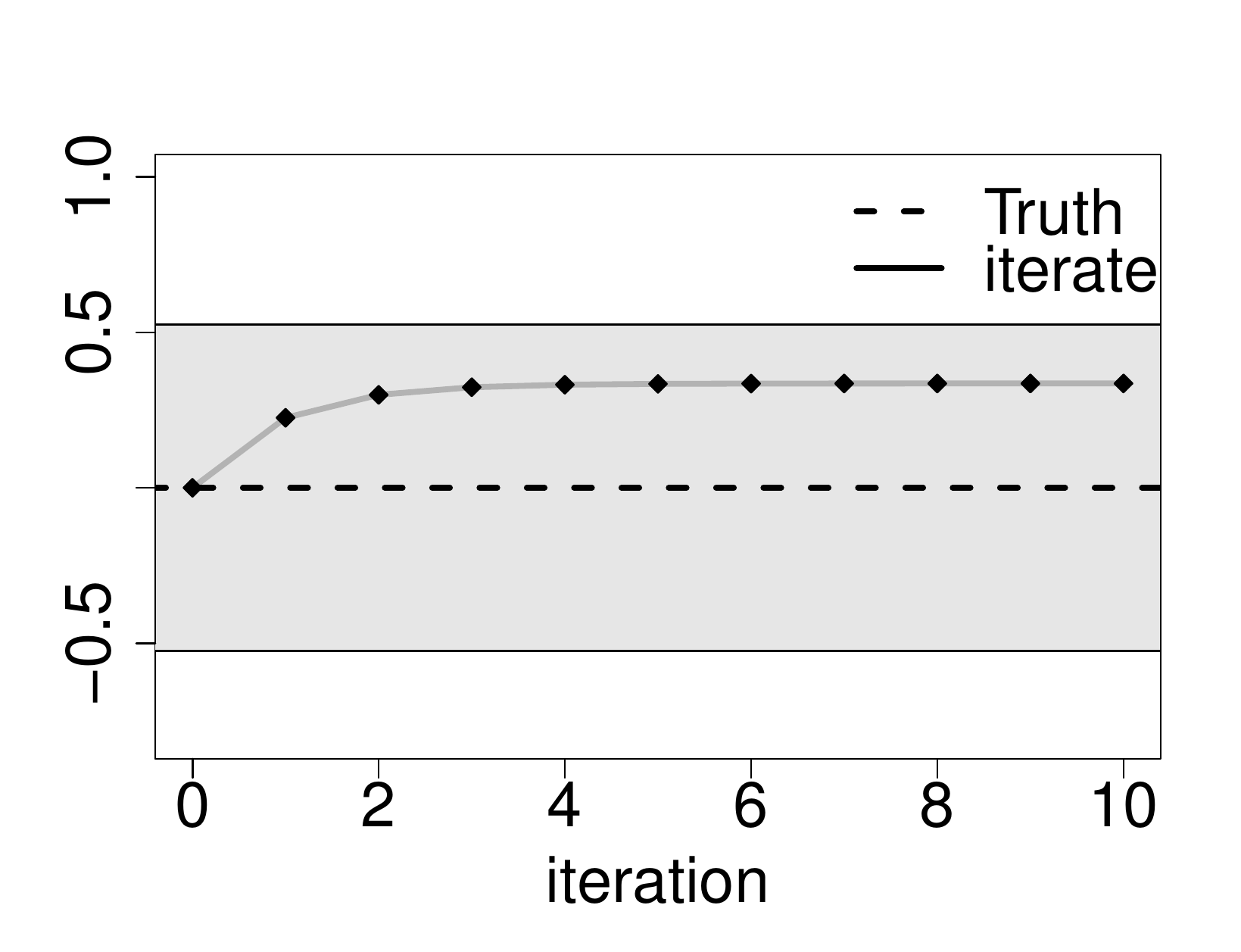}
\end{tabular}
\caption{An example of the realizations from the iterative algorithm in Algorithm.~\ref{Alg:BFLasso} with $T=10$ iterations and $(n,\,p)=(100,\,500)$ (under the same setting as the numerical experiment in Section~\ref{Sec:simulation} with a Toeplitz type random design). The left panel corresponds to the traceplot for a non-zero signal; and the right panel corresponds to a zero signal. They are both initialized at the unadjusted Lasso estimate---the first iteration is approximately the de-sparsified Lasso estimate, and the iterates converge to the CLasso estimate. The shaded region corresponds to a interval centered at the truth with the same length as a $95\%$ confidence interval. Therefore, any confidence interval centered at the point outside the shaded region will not cover the truth. Both figures illustrates that one iteration is not enough for $\theta^t$ to fully escape from the super-efficiency exhibited in the initial Lasso estimate, leading to under-coverages for non-zero signals and over-coverages for zero signals when confidence intervals are constructed based on the de-sparsified Lasso.}
\label{fig:2}
\end{figure}

\noindent According to Proposition~\ref{Prop:Relation}, the de-sparsified Lasso estimator $\widehat{b}_1$ is asymptotically equivalent to the first iterate $\theta^1$ in the iterative Algorithm.~\ref{Alg:BFLasso} when initialized at the original Lasso estimator. As a consequence, the de-sparsified Lasso estimator tends to be close to the Lasso estimator when the convergence of the algorithm is slow. Since the Lasso estimator has super-efficiency---meaning that it shrinks small signals to be exactly zero and incurs some amount of shrinkage for non-zero signals, the de-sparsified estimator may also inherit the super-efficiency from the Lasso to some extent. 
This explains our empirical observations in Section~\ref{Sec:simulation} that for the de-sparsified Lasso, the coverage probabilities of confidence intervals for unknown true signals with non-zero values tend to significantly below the nominal level, while the coverage probabilities of zero signals almost always attain, even exceed the nominal level. In comparison, the CLasso estimator, a refined de-sparsified Lasso estimator though applying more iterations, tends to be fully escaped from the local super-efficiency region. Consequently, the penalty-induced bias in those non-zero signals has been fully corrected and the shrinking-to-zero signals have been fully released from zero (see Figure.~\ref{fig:1} and Figure.~\ref{fig:2} and captions therein for an illustration). As a result, the CLasso tends to produce a more balanced coverage probabilities between zero and non-zero signals (see our empirical studies in Section~\ref{Sec:simulation} for more details). This improvement over the de-sparsified Lasso becomes more prominent as the convergence of the iterative algorithm becomes slow, that is, when the algorithmic convergence rate $\sqrt{(s^2/n)\log p}$ (see Theorem~\ref{Thm:convergence}) becomes relatively large.

\section{Theory}\label{Sec:Theory}
We show the asymptotic normality of the CLasso estimator under suitable conditions on $\alpha$ and the design matrix in Section~\ref{Sec:Normality}. In Section~\ref{Sec:SemiEff}, we show that the $\alpha$ constructed via node-wise regression~\eqref{Eqn:Choose_alpha} satisfies those conditions and in addition leads to asymptotic optimality in terms of semiparametric efficiency. In Section~\ref{Sec:Alg_rate}, we turn to the algorithmic aspect of the CLasso by showing a globally linear contraction rate of the iterative algorithm proposed in Section~\ref{Sec:Algorithm}.

\subsection{Asymptotic normality of the Constrained Lasso}\label{Sec:Normality}
For technical convenience, we study the following variant of the CLasso problem by adding constraint $\|\gamma\|_1\leq \rhobar$ for some sufficiently large $\rhobar$ so that the truth $\gammastar$ is feasible in the second Lasso programming,
\begin{subequations}\label{Eqn:M_est_const}
\begin{empheq}[left=\empheqlbrace]{align} 
&\frac{1}{n}\, (X-Z\alpha)^T(Y - X \theta-Z\gamma) = 0 \label{Eqn:Za_const}\\
&\gamma \in \argmin_{r\in\mb R^p,\,\|r\|_1\leq \rhobar} \Big\{\frac{1}{2n}\,\|Y-X\theta-Zr\|^2 + \lambda\,\|r\|_1\Big\}.\label{Eqn:Zb_const}
\end{empheq}
\end{subequations}
This additional constraint becomes redundant as long as the initialization $\gamma^0$ of the iterative algorithm satisfies $\|\gamma\|_1\leq \rhobar$, since our proof indicates that for any $t$, the time-$t$ iterate $\gamma^t$ still satisfies the same constraint.

Recall that the true data generating model is $Y=X\thetastar+Z\gammastar+w$ with $w\sim\m N(0,\,\sigma^2I_n)$ and $\gammastar$ is assumed to be $s$-sparse. It is good to keep in mind that we are always working in the regime that $d\ll n$ and $s\,\log p/\sqrt{n}\ll 1$. For any $m$ by $n$ matrix $A=(A_{ij})$, we denote its element-wise sup norm by $\|A\|_\infty=\max_{i,j}|A_{ij}|$, the $\ell_\infty$ to $\ell_\infty$ norm by $\|A\|_{\infty,\infty}=\max_{i}\|A^i\|_1$, the $\ell_1$ to $\ell_\infty$ norm by $\|A\|_{1,\infty}=\sum_{i}\|A_j\|_\infty$. Let $S$ denote the index set corresponding to the support of the $s$-sparse vector $\gammastar$. Let $\m C=\{a\in\mb R^p:\, \|a_{S^c}\|_1\leq 3\|a_S\|_1\}$ denote a cone in $\mb R^p$. This cone plays a key role in the analysis, since we will show that $\gammahat$ as well as any iterate $\gamma^t$ in Algorithm.~\ref{Alg:BFLasso} belongs to this cone with high probability due to the $\ell_1$ regularization. Recall that $\Xtil =X-Z\alpha$ is the residual of $X$ after the $Z$ part has been removed.

We make the following assumption on the design matrix $Z$ for the nuisance part, which is a standard assumption in high-dimensional linear regression under the sparsity constraint (for discussions about this condition, see, for example, \cite{bickel2009simultaneous}).
\paragraph{Restricted eigenvalue condition (REC):} The nuisance design matrix $Z$ satisfies
\begin{align*}
\inf_{u\in \m C}\,  \frac{1}{n}\,\frac{\|Zu\|_2^2}{\|u\|_2} \geq \mu.
\end{align*}

\begin{theorem}\label{Thm:Normality}
Assume REC. Moreover, suppose there are constants $(C,\,\tau,\,\nu)$ such that $\|(n^{-1}\Xtil^T\Xtil)^{-1}\|_{\infty} \leq C$, $\|(n^{-1}\Xtil^T\Xtil)^{-1}\|_{\infty,\infty} \leq C$, $\|n^{-1}\Xtil^TZ\|_\infty \leq \tau$, $\|n^{-1}\Xtil^TZ\alpha\|_\infty \leq \nu \leq (2Cd)^{-1}$, $\|n^{-1}\Xtil^T\Xtil\|_2 \leq C$, and the design matrices $(X,Z)$ has been normalized so that $\max_{j=1,\ldots,d}\|n^{-1/2}X_j\|^2\leq C$ and $\max_{j=1,\ldots,p}\|n^{-1/2}Z_j\|^2\leq C$. If $\displaystyle \lambda \geq 2\sigma\,\sqrt{\frac{2C\log p}{n}} +\frac{4\sigma\,C^2d}{\sqrt{n}}+ 8Cd\,\rhobar\,\tau$ and the truth $\gammastar$ of the nuisance parameter satisfies $\|\gammastar\|_1\leq \rhobar$ and is $s$-sparse, then for some constant $c>0$,
\begin{align*}
\sqrt{n}\,(\thetahat-\thetastar) &= W + \Deltahat,\\
W = \sqrt{n}\,(\Xtil^T\Xtil)^{-1} \Xtil^Tw &\sim \m N\big(0,\, \sigma^2 (n^{-1}\Xtil^T\Xtil)^{-1}\big),\\
\mb P\bigg[\|\Deltahat\|_\infty\geq \frac{6\sqrt{n}\, \tau\,s\,\lambda}{\mu} + 2C^2\,&\sigma\,\nu + 4C\,\sqrt{n}\,\rhobar\,\tau\,\nu\bigg] \leq p^{-c}+n^{-c},
\end{align*}
where the randomness is with respect to the noise vector $w$ in the linear model.
\end{theorem}
\noindent 
Theorem~\ref{Thm:Normality} shows that if the remainder term $\Deltahat=o_P(1)$, then $\sqrt{n}\,(\thetahat-\thetastar)$ is asymptotically equivalent to a normally distributed vector $W$. This theorem applies to any design $(X,\,Z)$ and matrix $\alpha$, and does not use any randomness in them.
Let us make some quick remark regarding the conditions in Theorem~\ref{Thm:Normality}. Since we are interested in the regime that $d\ll n$, or more simply, $d=1$, assumptions on $\Xtil$ like $\|(n^{-1}\Xtil^T\Xtil)^{-1}\|_{\infty} \leq C$, $\|(n^{-1}\Xtil^T\Xtil)^{-1}\|_{\infty,\infty} \leq C$ and $\|n^{-1}\Xtil^T\Xtil\|_2 \leq C$ are easily satisfied for some sufficiently large $C$ (see, for example, Theorem~\ref{Thm:alpha}). The design matrix column normalization condition is also standard. The less obvious assumptions are $\|n^{-1}\Xtil^TZ\|_\infty \leq \tau$ and $\|n^{-1}\Xtil^TZ\alpha\|_\infty \leq \nu \leq (2Cd)^{-1}$, which controls the bias magnitude in $\thetahat$ and critically depends on the choice of $\alpha$. More importantly, in order to make the remainder term $\Deltahat$ in the local expansion of $\thetahat$ to vanish, $(\tau,\nu)$ needs to decay reasonably fast as $n\to 0$. In Theorem~\ref{Thm:alpha} below, we show that under mild assumptions on the design, the $\alpha$ constructed via node-wise regression~\eqref{Eqn:Choose_alpha} has nice properties that makes $\tau \leq C'\,\sqrt{n^{-1}\log p}$ and $\nu \leq C'\,\sqrt{n^{-1}\log p}$ hold with high probability with respect to the randomness in the design. By plugging in these bounds, Theorem~\ref{Thm:Normality} implies that remainder term $\Deltahat=O_P\big((s/\sqrt{n})\,\log p\big)=o_P(1)$ is indeed of higher-order relative to $W=O_P(1)$ as $(s/\sqrt{n})\,\log p\to 0$ and $n\to \infty$. Although we assume the noise $w$ in the linear model to be Gaussian, the proof can be readily extend to noises with sub-Gaussian tails.

\subsection{Semiparametric efficiency of the CLasso}\label{Sec:SemiEff}
In this subsection, we show that the matrix $\alpha$ chosen via optimization procedure~\eqref{Eqn:Choose_alpha} satisfies the conditions in Theorem~\ref{Thm:Normality}. Moreover, the corresponding CLasso estimator $\thetahat$ is semiparametric efficient---it has the smallest asymptotic variance, or achieves the Cram\'{e}r-Rao lower bound from a semiparametric efficiency perspective. Recall that $\Xtil=X-Z\alpha$ is the residual matrix, where $\alpha$ is the solution of the node-wise regression~\eqref{Eqn:Choose_alpha}. Let $\Omega^\ast=\big(\mb E[(X^i-Z^i\alphastar)^T(X^i-Z^i\alphastar)]\big)^{-1}$ denote the inverse of the semiparametric efficient information matrix of $\theta$, which is also the Cram\'{e}r-Rao lower bound of the asymptotic covariance matrix of any asymptotically unbiased estimator of $\theta$ (see \cite{van2014} for more details on the precise definition of semiparametric optimality of $\theta$).
Recall that we assume both design matrices $X$ and $Z$ to be random.

\paragraph{Assumption D:} Let $U=(X, \,Z)\in\mb R^{n\times (d+p)}$ denote the entire design matrix. Rows $\{U^i\}_{i=1}^n$ of $U$ are i.i.d.~with zero mean and sub-Gaussian tails, that is, $\mb E[U^i]=0$ and there exists some constant $C_0$, such that for any vector $h\in\mb R^{d+p}$, 
\begin{align*}
\mb E\big[\exp\{U^ih\}\big] \leq \exp\Big\{\frac{C_0}{2}\, \|h\|^2\Big\}.
\end{align*}

\begin{theorem}\label{Thm:alpha}
If Assumption D holds and $\lambda_j \geq 2C_0\sqrt{n^{-1}\log p}$, then in the node-wise regression~\eqref{Eqn:Choose_alpha}, with probability at least $1-d\,p^{-c}$ with respect to the randomness in the design $(X,\,Z)$, we have
\begin{align*}
\|n^{-1}\Xtil^TZ\|_\infty \leq \max_j\lambda_j\quad\mbox{and}\quad \|n^{-1}\Xtil^TZ\alpha\|_\infty \leq 3\,\max_j \lambda_j \,\max_j\|\alphastar_j\|_1.
\end{align*}
In addition, if we choose $\lambda_j = 2C_0\,D\,\sqrt{n^{-1}\log p}$, then for some constant $C'$ depending on $D$, the largest eigenvalue of $\Omega^\ast$ and $\alphastar$, it holds with probability at least $1-d\,p^{-c}-d^2\,n^{-c}$ with respect to the randomness in the design that
\begin{align*}
\bigg\|\Big(\frac{\Xtil^T\Xtil}{n}\Big)^{-1} -\Omega^\ast \bigg\|_\infty \leq C'\, \sqrt{\frac{\log p}{n}}.
\end{align*}
\end{theorem}
\noindent The choice of $\lambda_j$ heavily depends on the tail behavior of the design $U$. For example, if the design instead has a heavier sub-exponential tail, then we need to increase the regularization parameter to $\Omega(\log p /\sqrt{n})$. Similar to the theory in \cite{javanmard2014confidence}, we do not need to impose any sparsity condition on $\alphastar_j$'s as in \cite{van2014}---the only assumption is the boundedness of $\max_{j}\|\alphastar_j\|_1$, which tends to be mild and satisfied in most real situations. In fact, in the proof we find that a ``slow rate" type bound \cite{buhlmann2011statistics} for the $\ell_1$ penalized estimator suffices for the proof and we do not need to go to the ``fast rate" regime that demands sparsity. 

Theorem~\ref{Thm:alpha} also implies that we may choose the critical quantities $\tau$ and $\nu$ appearing in Theorem~\ref{Thm:Normality} to be of order $\sqrt{n^{-1}\log p}$. Finally, by combining Theorem~\ref{Thm:Normality} and Theorem~\ref{Thm:alpha} with Slutsky's theorem, we obtain the following corollary showing the semiparametric optimality of  the CLasso estimator $\thetahat$.
\begin{corollary}\label{Cor:SemiOpt}
Under the assumptions in Theorem~\ref{Thm:Normality} and Theorem~\ref{Thm:alpha}, we have that as $n\to\infty$ and $\displaystyle \frac{s\log p}{\sqrt{n}} \to 0$,
\begin{align*}
\sqrt{n}\,(\thetahat-\thetastar)\overset{d}{\rightarrow} \m N\big(0,\, \sigma^2\, \Omega^\ast\big).
\end{align*}
\end{corollary}

\subsection{Confidence intervals and hypothesis testing}
In this subsection, we construct asymptotically valid statistical inference procedures based on the form of the asymptotic normal limit of $\thetahat$. 

\paragraph{Confidence intervals:} For any $d$-dimensional vector $r$, we can construct an $(1-\alpha)$ confidence interval for linear functional $r^T\theta$ as
\begin{align}\label{Eqn:CIr}
J_r(\alpha) =\Big[r^T\thetahat - \frac{z_{\alpha/2}\,\widehat{\sigma}}{\sqrt{r^T(\Xtil^T\Xtil)^{-1}r}},\,r^T\thetahat + \frac{z_{\alpha/2}\,\widehat{\sigma}}{\sqrt{r^T(\Xtil^T\Xtil)^{-1}r}}\Big],
\end{align}
where $z_{\alpha/2}$ denotes the $1-\alpha/2$ quantile of a standard normal distribution, and $\widehat{\sigma}$ is any consistent estimator of the noise level $\sigma$, for example, the scaled Lasso estimator \cite{sun2012scaled},
\begin{align*}
(\widehat{\beta}_{S},\,\widehat{\sigma}_S):\,=\argmin_{\beta\in\mb R^{d+p},\sigma>0}\Big\{\frac{1}{2n\sigma}\,\|Y-U\beta\|^2 +\frac{\sigma}{2}+\widetilde{\lambda}\,\|\beta\|_1\Big\},
\end{align*}
with the universal penalty $\widetilde{\lambda}=\sqrt{(2/n)\log p}$. Theorem~\ref{Thm:alpha} and Corollary~\ref{Cor:SemiOpt} combined with Slutsky's theorem imply
\begin{align*}
\frac{\widehat{\sigma}}{\sqrt{r^T(\Xtil^T\Xtil)^{-1}r}}\,\big(r^T\thetahat-r^T\thetastar\big) \overset{d}{\rightarrow} \m N\big(0,\, I_d\big), \quad\mbox{as $n\to\infty$ and $(s/\sqrt{n})\,\log p \to 0$,}
\end{align*}
where we use notation $Q^{1/2}$ to denote the square root for any symmetric matrix $Q$.
Consequently, we have for any $\alpha\in(0,1)$,
\begin{align*}
\mb P\big[r^T\thetastar \in J_r(\alpha)\big]=\mb P\Big[\Big|\frac{\widehat{\sigma}}{\sqrt{r^T(\Xtil^T\Xtil)^{-1}r}}\,\big(r^T\thetahat-r^T\thetastar\big)\Big| \leq z_{\alpha/2}\Big]
\to \alpha, \ \ \mbox{as $n\to\infty$ and $(s/\sqrt{n})\,\log p  \to 0$,}
\end{align*}
implying that $J_r(\alpha)$ is an asymptotically valid confidence interval with significance level $1-\alpha$ for $r^T\theta$. 

In the special case when we are only interested in one component of $\beta$, say $\beta_j$, in the linear model~\eqref{Eqn:LM}, then in order to minimize the asymptotic length of its confidence interval, we take $\theta=\beta_j$ as the parameter of interest and $\gamma=\beta_{-j}$ as the nuisance parameter in the semiparametric formulation~\eqref{Eqn:Linear_Model}, where for any vector $a$ we use notation $a_{-j}$ to denote the its sub-vector without the $j$-th component. Let $X_j$ and $Z_{-j}$ to denote the corresponding design matrices. Then, the previous procedure leads to an asymptotically valid $(1-\alpha)$-confidence interval of $\beta_j$ as
\begin{align}\label{Eqn:SCI}
\Big[ \widehat{\beta}_j - \frac{z_{1-\alpha/2}\,\widehat{\sigma}}{\|X_j - Z_{-j}\alpha_j\|},\, \widehat{\beta}_j + \frac{z_{1-\alpha/2}\,\widehat{\sigma}}{\|X_j - Z_{-j}\alpha_j\|}\Big].
\end{align}

\paragraph{Hypothesis testing:} By converting the confidence interval~\eqref{Eqn:CIr}, we can construct the following asymptotically valid procedure for testing $H_0:\, r^T\theta=u$ vs $H_1:\, r^T\theta\neq u$ for any contrast $r^T\theta$ by rejecting $H_0$ if
\begin{align*}
\Big|\frac{\widehat{\sigma}}{\sqrt{r^T(\Xtil^T\Xtil)^{-1}r}}\,\big(r^T\thetahat-u\big)\Big| \geq z_{\alpha/2}.
\end{align*}
By a similar argument, it can be shown that this testing procedure has an asymptotic type I error $\alpha$. By converting the individual confidence intervals~\eqref{Eqn:SCI}, we can construct $p$-values for each $\beta_j$ as
\begin{align}\label{Eqn:Pvalues}
P_j = 2\, \Big( 1- \Phi\Big(\frac{|\widehat{\beta}_j|\,\|X_j-Z_{-j}\alpha_j\|}{\widehat{\sigma}}\Big)\Big), \quad j=1,2,\ldots,p,
\end{align}
where $\Phi$ denotes the cdf of the standard normal distribution.
We may use the Bonferroni--Holm procedure to control the asymptotic family-wise error rate (FWER) to be within $\alpha$ for multiple testing $H^j_0:\,\beta_j=0$ vs $H^j_1:\,\beta_j\neq 0$. More specifically, we first sort the $p$ p-values as $P_{(1)},P_{(2)},\ldots,P_{(p)}$, whose associated hypotheses are $H^{(1)},H^{(2)},\ldots,H^{(p)}$; then find the minimum index $k$ such that $P_{(k)}>\alpha/(p+1-k)$ (if $k$ does not exist, then set $k=p+1$), and reject the hypotheses $H^{(1)},\,\ldots,\, H^{(k-1)}$ if $k >1$.

\subsection{Convergence analysis of the iterative algorithm}\label{Sec:Alg_rate}
In this subsection, we characterize the convergence of the iterative algorithm described in Section~\ref{Sec:Algorithm} for solving CLasso.

\begin{theorem}\label{Thm:convergence}
Suppose the assumptions of Theorem~\ref{Thm:Normality} holds. If the regularization parameter satisfies $\displaystyle \lambda_t = D\,\Big\{2\sigma\,\sqrt{\frac{2C\log p}{n}} +\frac{4\sigma\,C^2d}{\sqrt{n}}+ 8Cd\,\tau\, \|\gamma^{t-1}-\gammastar\|_1\Big\}$ for some $D\geq 1$ and $48\, CD\, s\,\tau\,\mu^{-1}<1$, then with probability at least $1-d\,p^{-c}-d^2\,n^{-c}$,
\begin{align*}
&\|\sqrt{n}\,(\theta^t-\thetastar) - W\|_\infty \leq 3\sqrt{n}\,\tau\,\rho^{t-1}\,\|\gamma^0-\gammastar\|_1 + \vep_n, \quad \forall t\geq1, \\
\mbox{where}\quad &\rho=48\, CD\, \frac{s\,\tau}{\mu} \quad\mbox{and}\quad\vep_n = 36\,CD\,\frac{\sigma\,\tau\,s}{(1-\rho)\,\mu} \,\sqrt{2C\,\log p}+ 72\,C^2D\,\frac{\sigma\,\tau\,d}{1-\rho}+2\,C^2\, \sigma\,\nu,
\end{align*}
where $W$ is defined in Theorem~\ref{Thm:Normality}.
\end{theorem}
\noindent This theorem shows that our iterative algorithm enjoys globally linear convergence up to the statistical precision of the model, meaning the typical distance between the rescaled estimator $\sqrt{n}\,(\theta^t-\thetastar)$ and its non-degenerate asymptotic normal limit.

As we mentioned in Section~\ref{Sec:Algorithm}, it would be beneficial to consider a sequence of decreasing regularization parameters $\{\lambda_t:\,t\geq 1\}$. Now we provide a formal explanation. In fact, a smaller $\lambda_t$ leads to a smaller bias in $\gamma^t$, which will in turn reduce the higher-order error $\Deltahat$ in Theorem~\ref{Thm:Normality} (by identifying $\rhobar$ with $\|\gamma^{t-1}-\gammastar\|_1$) and improves the accuracy of the normal approximation to $\sqrt{n}\,(\thetahat-\thetastar)$. However, at the beginning of the algorithm where initialization $\gamma^0$ may be far away from $\gammastar$,
we need a large $\lambda^t$ to enforce the algorithm to converge. Therefore, at least theoretically, by adopting a sequence of decreasing $\lambda_t$'s we can achieve both globally linear convergence of the algorithm as well as accurate normal approximation to the final estimator.
A combination of the previous results with Theorem~\ref{Thm:convergence} leads to the following corollary characterizing the algorithmic rate of convergence in terms of the difficulty of the problem reflected by $(s,\,n,\,p)$.

\begin{corollary}\label{Cor:AlgRate}
Under the assumptions in Theorem~\ref{Thm:Normality}, Theorem~\ref{Thm:alpha} and Theorem~\ref{Thm:convergence}, there exists some constants $(c_0,\,c_1,\,c_2,\,c_3)$ independent of $(s,\,n,\,p)$, such that with probability at least $1-d\,p^{-c}-d^2\,n^{-c}$, 
\begin{align}\label{Eqn:AlgRate}
&\|\sqrt{n}\,(\theta^t-\thetastar)- W\|_\infty \leq c_1\,\sqrt{\log p}\,\Big(c_2\,\frac{s^2\,\log p}{n}\Big)^{\frac{t-1}{2}}\,\|\gamma^0-\gammastar\|_1 + c_3\,\frac{s\,\log p}{\sqrt{n}},\quad \forall t\geq1.
\end{align}
\end{corollary}
\noindent 
Different from gradient-based procedures where the optimization error typically contracts at a constant factor independent of sample size $n$ and dimensionality $p$ (depends on the conditional number) of the problem, Corollary~\ref{Cor:AlgRate} shows that the proposed iterative algorithm exhibits a contraction factor proportional to $\sqrt{(s^2/n)\,\log p}$ that decays towards zero as $(s^2/n)\,\log p \to 0$. Therefore, the proposed algorithm lies in between first-order based gradient methods and second-order based Newton's methods (however, the comparison between gradient method and our algorithm may not fully fair since we have ignored the computational complexity in solving the Lasso programming).

If we initialize the algorithm at the Lasso estimate as in Algorithm.~\ref{Alg:BFLasso}, then $\|\gamma^0-\gammastar\|_1\sim s\,\sqrt{n^{-1}\log p}$. At the first iteration $t=1$ (which corresponds to the de-sparsified Lasso estimate, see Proposition~\ref{Prop:Relation} for a precise statement), the first term on the right hand side of bound~\eqref{Eqn:AlgRate} has the same order $(s/\sqrt{n})\,\log p$ as the second term. As a consequence, although the de-sparsified Lasso estimator achieves the same asymptotic error rate towards a normal limit as the CLasso estimate, the latter still has the potential to reduce the constant in front of the rate through applying more iterations. In our numerical experiments in the next section,  we empirically illustrate that the gain in terms of reducing the constant can be prominent.

\section{Empirical results}\label{Sec:simulation}
In this section, we first compare the CLasso with the de-sparsified Lasso via simulations and then apply the CLasso to a real dataset.

\subsection{Synthetic data}
We generate the synthetic dataset from the following linear model (matrix form)
\begin{align*}
Y = X\,\theta +w,\quad w\sim\m N(0,\,I_n),
\end{align*}
where $X\in\mb R^{n\times p}$ is the design matrix, $\theta=(\theta_1,\,\theta_2,\ldots,\,\theta_p)^T\in\mb R^p$ is the unknown regression coefficient vector and the noise has unit variance. We consider different combinations between sample size $n\in\{100,\,500,\,1000\}$ and dimensionality $p\in\{100,\,500\}$.  Suppose that we are interested in the 3rd and 7th components $(\theta_3,\,\theta_7)$ of $\theta$. By considering $\theta_j$ as the one-dimensional parameter of interest, we rewrite the model as
\begin{align}\label{Eqn:Individual}
Y = X_j\, \theta_j+Z_j \gamma_j +w,\quad w\sim\m N(0,\,I_n),
\end{align}
where $Z_j=X_{-j}$, the sub-matrix of $X\in\mb R^{n\times p}$ with the $j$th column being removed, is the nuisance design matrix, and $\gamma_j=\theta_{-j}$, the sub-vector of $\theta\in\mb R^p$ without the $j$th element, is the nuisance parameter. We run the CLasso for $j=3$ and $j=7$, respectively, and construct confidence intervals for $\theta_3$ and $\theta_7$. Here, we do not treat $(\theta_3,\,\theta_7)^T$ as a two-dimensional parameter of interest and construct individual confidence intervals based on their joint asymptotic normal distribution, since this leads to increased lengths for the individual confidence intervals and decreased powers for the individual hypothesis testing procedures.

In the linear model, the true regression coefficient vector is set to be
$$\thetastar=\big(2,\, -1,\, -2,\, 3,\, 1,\, 0,\ldots,\, 0\big)^T\in\mb R^p,$$ 
so that $X_3$ is an relevant predictor with none-zero signal strength, and $X_7$ is an unimportant predictor with zero signal strength, and the overall sparsity level is $s=5$.
The rows of $X\in\mb R^{p}$ are i.i.d.~realizations from $\m N_p(0,\, \Sigma)$. We consider two types of $\Sigma$:
\begin{align*}
&\mbox{Toeplitz:} \qquad\ \   \Sigma_{jk} = 0.9^{|j-k|},\\
&\mbox{Equi corr:} \qquad \Sigma_{jk}=0.8 \quad\mbox{for }j\neq k,\quad \Sigma_{jj}=1\quad \mbox{for all $j$.}
\end{align*}

\begin{table}[tbh]
  \centering
  \setlength{\tabcolsep}{0.7em} 
{\renewcommand{\arraystretch}{1.2}
\begin{tabular}{cccccccc}
 \hline\hline
  &  & \multicolumn{3}{c}{{\bf Toeplitz}} &  \multicolumn{3}{c}{{\bf Equi corr}}\\
 {\bf Measure} & {\bf Method} & $n=100$ & $n=500$ & $n=1000$ & $n=100$ & $n=500$ & $n=1000$ \\
    \hline 
\multirow{3}{*}{Cov $\theta_3$}    & UP Lasso & 0.00 & 0.01  & 0.07 & 0.08 & 0.11 &   0.13 \\ 
 & CLasso & 0.94 & 0.96 & 0.94 & 0.91 & 0.95 &  0.96 \\ 
& DS Lasso & 0.06 & 0.34 & 0.64 & 0.25 & 0.73 &  0.82 \\ \hline
\multirow{3}{*}{Error  $\theta_3$}    & UP Lasso & 1.861 & 0.738 & 0.376 & 0.968 & 0.310 &  0.225 \\ 
  & CLasso & 0.328& 0.136 &  0.106 & 0.298 &  0.098 & 0.069   \\
& DS Lasso &  0.920 &  0.376 & 0.190 & 0.711 & 0.171 &  0.106 \\ \hline
\multirow{3}{*}{Cov $\theta_7$}    & UP Lasso & 0.41 & 0.28 & 0.88 & 0.65 & 0.85 &  0.85 \\ 
 & CLasso & 0.88 & 0.93 & 0.95 & 0.91 & 0.94 & 0.95   \\ 
& DS Lasso & 0.96 &  0.97 & 0.97 & 0.95 & 0.97 & 0.97  \\ \hline
\multirow{3}{*}{Error  $\theta_7$}    & UP Lasso & 0.760 & 0.348 & 0.232 & 0.487 & 0.146 & 0.093  \\ 
 & CLasso & 0.427 & 0.116 & 0.110 & 0.332 & 0.109 & 0.073  \\
& DS Lasso & 0.297 & 0.097 & 0.096 & 0.252 & 0.092 &  0.064 \\
       \hline\hline
\end{tabular}
}
\caption{Confidence intervals in linear model with dimension $p=100$. Cov $\theta_3$ and Error $\theta_3$ are the coverage probability (significance level $0.95$) and the root mean squared error $\sqrt{\mb E\big[|\thetahat_3-\theta^\ast_3|^2\big]}$ of the non-zero signal $\theta_3$;  Cov $\theta_7$ and Error $\theta_7$ are for the zero signal $\theta_7$. UP Lasso is the naive un-penalized Lasso estimator described in \eqref{Eqn:UPLASSO}, CLasso is the proposed method, and DS Lasso is the de-sparsified Lasso proposed in \cite{van2014}. All numbers are based on average over $500$ replicates.} \label{table:1}
\end{table}

\begin{table}[tbh]
  \centering
  \setlength{\tabcolsep}{0.7em} 
{\renewcommand{\arraystretch}{1.2}
\begin{tabular}{cccccccc}
 \hline\hline
  &  & \multicolumn{3}{c}{{\bf Toeplitz}} &  \multicolumn{3}{c}{{\bf Equi corr}}\\
 {\bf Measure} & {\bf Method} & $n=100$ & $n=500$ & $n=1000$ & $n=100$ & $n=500$ & $n=1000$ \\
    \hline 
\multirow{3}{*}{Cov $\theta_3$}    & UP Lasso & 0.00 & 0.00  & 0.00 & 0.01 & 0.12 &   0.07 \\ 
 & CLasso & 0.89 & 0.95 & 0.95 & 0.87 & 0.93 &  0.94 \\ 
& DS Lasso & 0.03 & 0.08 & 0.37 & 0.14 & 0.63 &  0.73 \\ \hline
\multirow{3}{*}{Error  $\theta_3$}    & UP Lasso & 2.142 & 0.889 & 0.511 & 1.184 & 0.371 &  0.253 \\ 
  & CLasso & 0.383 & 0.131 &  0.104 & 0.345 &  0.114 & 0.081   \\
& DS Lasso &  1.035 &  0.472 & 0.248 & 0.842 & 0.204 &  0.124 \\ \hline
\multirow{3}{*}{Cov $\theta_7$}    & UP Lasso & 0.23 & 0.12 & 0.16 & 0.59 & 0.83 &  0.81 \\ 
 & CLasso & 0.89 & 0.92 & 0.94 & 0.90 & 0.93 & 0.95   \\ 
& DS Lasso & 0.98 &  0.96 & 0.97 & 0.95 & 0.96 & 0.97  \\ \hline
\multirow{3}{*}{Error  $\theta_7$}    & UP Lasso & 0.845 & 0.395 & 0.265 & 0.570 & 0.156 & 0.110  \\ 
 & CLasso & 0.417 & 0.166 & 0.108 & 0.369 & 0.120 & 0.080  \\
& DS Lasso & 0.267 & 0.134 & 0.093 & 0.242 & 0.100 &  0.070 \\ 
       \hline\hline
\end{tabular}  
}
\caption{Confidence intervals in linear model with dimension $p=500$. For details, see the caption of Table.~\ref{table:1}.} \label{table:2}
\end{table}

We compare the CLasso with the un-penalized Lasso (UP Lasso) in~\eqref{Eqn:UPLASSO} and the de-sparsified Lasso (DS Lasso) proposed in \cite{van2014}. Note that the UP Lasso can also be implemented via Algorithm.\ref{Alg:BFLasso} by setting $\alpha\equiv 0$ throughout. We use the scaled Lasso \cite{sun2012scaled} with its universal regularization parameter to find an estimate $\widehat{\sigma}^2$ of the error variance, and then set $\lambda=\widehat{\sigma}\sqrt{(2\,\log p)/n}$ as the regularization parameter in all three methods (the same procedure is applied for setting the regularization parameters in the node-wise regression~\eqref{Eqn:Choose_alpha} for finding $\alpha$). We use the R package \texttt{glmnet} \cite{friedman2010regularization} to fit the Lasso programming for updating $\gamma$ in Algorithm.\ref{Alg:BFLasso}. In the CLasso, we construct $95\%$ confidence interval for $\theta_j$ ($j=3$ and $7$) via~\eqref{Eqn:SCI} and compute the p-values via~\ref{Eqn:Pvalues}.

Table.~\ref{table:1} and Table.~\ref{table:2} report the root mean square errors and coverages of $95\%$ confidence intervals under $p=100$ and $p=500$, respectively. We record the empirical coverage frequency over $500$ replicates in each combination of $(n,\, p)$ and the mean square error of estimating the parameters $\theta_{j}$ ($j=3,\,7$). In all scenarios, the UP Lasso has poor performance as we may expect, suggesting that a properly chosen $\alpha$ matrix is critical for the CLasso to work.
As expected, the coverage probability for the non-zero signal $\theta_3$ via the DS Lasso is always lower than its nominal level $0.95$, accompanied with significantly larger estimation error than the CLasso. For example, as the dimension $p$ grows from $100$ to $500$, the coverage of the DS Lasso decreases from $0.34$ to $0.08$ for the Toeplitz design under $n=500$, and the estimation error are on average $2$ times larger than that of the CLasso. In contrast, the coverage of the CLasso for $\theta_3$ fluctuates around its nominal level $0.95$ when the dimension $p=100$, and in the much harder $p=500$ case, it steadily grows towards $0.95$ as the sample size goes from $100$ to $1000$. For the zero-signal $\theta_7$, the DS Lasso tends to have over-coverage, meaning that the coverage probability tends to exceed $0.95$ by an noticeable amount, which is also consistent with our theory presented in Section~\ref{Sec:Relation}. In comparison, the CLasso exhibits balanced coverage probabilities for both zero and non-zero signals---for both signals, the coverage probabilities are around the nominal level $0.95$. Again, as we can expect, because the DS Lasso estimator has not fully escaped from the super-efficiency behaviour of the Lasso estimator (see Section~\ref{Sec:Relation}), the estimation errors of the DS Lasso for the zero signal $\theta_7$ are consistently smaller than that  of the CLasso, even though the latter also achieves the nominal coverage probability of the confidence intervals.

\begin{table}[tbh]
  \centering
  \setlength{\tabcolsep}{0.7em} 
{\renewcommand{\arraystretch}{1.2}
\begin{tabular}{cccccccc}
 \hline\hline
  &  & \multicolumn{3}{c}{{\bf Toeplitz}} &  \multicolumn{3}{c}{{\bf Equi corr}}\\
 {\bf Measure} & {\bf Method} & $n=100$ & $n=500$ & $n=1000$ & $n=100$ & $n=500$ & $n=1000$ \\
    \hline 
\multirow{3}{*}{Power}    & UP Lasso & 0.47 & 1.00  & 1.00 & 0.68 & 1.00 &   1.00 \\ 
 & CLasso & 0.63 & 1.00 & 1.00 & 0.74 & 1.00 &  1.00 \\ 
& DS Lasso & 0.40 & 1.00 & 1.00 & 0.65 & 1.00 &  1.00 \\ \hline
\multirow{3}{*}{FWER}    & UP Lasso & 0.06 & 0.04 & 0.00 & 0.08 & 0.00 &  0.00 \\ 
  & CLasso & 0.04 & 0.00 &  0.00 & 0.02 &  0.00 & 0.00   \\
& DS Lasso &  0.00 &  0.00 & 0.00 & 0.00 & 0.00 &  0.00 \\ 
       \hline\hline
\end{tabular}  
}
\caption{Multiple testing in linear model with dimension $p=500$ with nominal FWER equal to $0.05$. UP Lasso is the naive un-penalized Lasso estimator described in \eqref{Eqn:UPLASSO}, CLasso is the proposed method, and DS Lasso is the de-sparsified Lasso proposed in \cite{van2014}. All numbers are based on average over $200$ replicates.} \label{table:3}
\end{table}

Table.~\ref{table:3} reports the average powers and FWERs for multiple testing $H^j_0:\,\beta_j=0$ vs $H^j_1:\,\beta_j\neq 0$, $j=1,\ldots,p$ under $p=500$. We use the Bonferroni--Holm (BH) procedure to control the FWER to be within $0.05$. The average power is defined as the empirical version of 
\begin{align*}
\mbox{Power}=5^{-1}\,\sum_{j=1}^5\mb P\big[H^j_0\mbox{ is rejected}\big],
\end{align*} 
and the average FWER the empirical version of
\begin{align*}
\mbox{FWER}=\mb P\big[\mbox{for at least one $j\geq 6$, $H^j_0$ is rejected}\big].
\end{align*}
Note that in the true data generating model, $(\beta_1, \beta_3, \beta_5)$ have relatively large signal to noise levels, explaining that most powers are around $0.6$ when sample size $n$ is small. According Table.~\ref{table:3}, the UP Lasso seems to have slightly better power than the DS Lasso, while the FWER of UP Lasso is worse (the BH procedure is only slightly less conservative than the Bonferroni correction, so a $0.06$ FWER based on the BH is pretty high). 
In contrast, the CLasso has the best power among the three at $n=100$ with a reasonably large FWER. As expected, the DS Lasso always has FWER close to zero because of the super-efficiency at zero inherited from the Lasso.
At $n=500$ and $1000$, all methods have power one and FWER close to zero due to the large sample size. 

\subsection{Real data application}

\begin{figure}[htb]
\centering
    \includegraphics[width=5in]{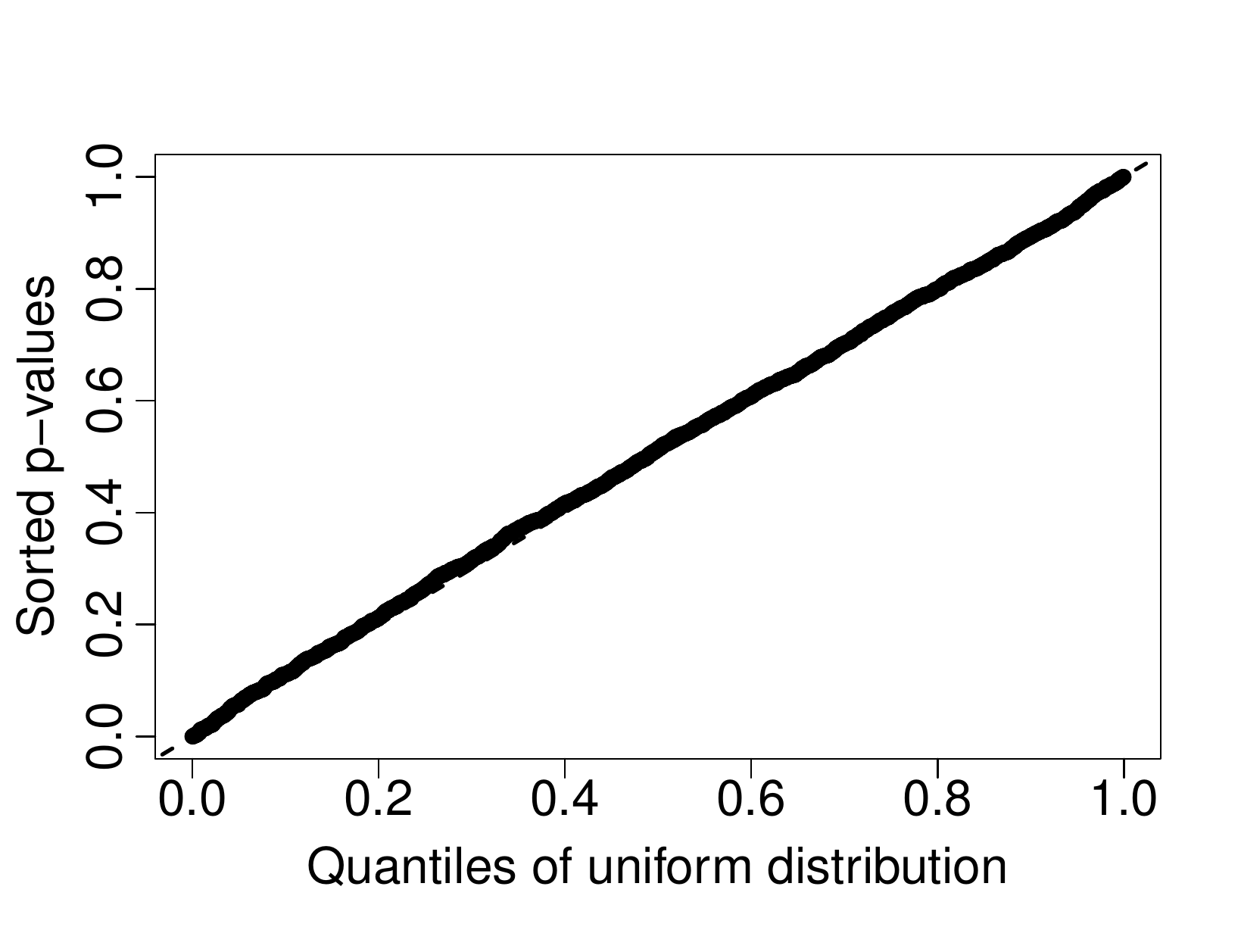}
\caption{Comparison between the empirical distribution of $p = 4,088$ computed p-values in the riboflavin example and the uniform distribution over $[0,\,1]$. The plot shows
that the p-values are distributed very closely to the uniform distribution. }
\label{fig:3}
\end{figure}

In this subsection, we apply the CLasso method to the riboflavin (vitamin B2)  production rate dataset. This data set is publicly available \cite{buhlmann2014high} and contains $n = 71$ samples and $p = 4,088$ covariates corresponding to the logarithm of the expression level of $4,088$ genes. The response variable for each sample is a real number indicating the logarithm of the riboflavin production rate. The same dataset has also been analyzed in \cite{van2014} and \cite{javanmard2014confidence}. Following \cite{van2014}, we model the data with a high-dimensional linear model and conduct individual hypothesis testing $H_0:\,\beta_j=0$ vs $H_1:\,\beta_j\neq 0$ for each gene via the semiparametric representation~\eqref{Eqn:Individual}. We find the $p$-value $P_j$ via~\eqref{Eqn:Pvalues}. The implementation of the CLasso is the same as in the synthetic data example. Figure.~\ref{fig:3} shows the empirical $p$-values computed from the data. The empirical distribution of the $p$-values follows a uniform distribution over $[0,\,1]$ reasonably well. After controlling the FWER to be within $5\%$ via the Bonferroni--Holm procedure, we find no significant regression coefficient, which is consistent with the conclusion drawn in \cite{van2014}, since the gene expressions are highly correlated and the number of covariates significantly exceeds the sample size ($\sqrt{n^{-1}\,\log p}\approx 0.34$).

\section{Proofs}\label{Sec:Proof}
In this section, we provide proofs of the main results in the paper.

\subsection{Proof of Theorem~\ref{Thm:Uniqueness}}
We apply the following lemma that shows any solution $\gamma$ of equation~\eqref{Eqn:M_est} is at most $C'\,s$ sparse for some constant $C'>0$ independent of $(n,\,p)$ and recall that $s$ is the sparsity level of the true unknown $\gammastar$. Let $\|\cdot\|_0$ denote the $\ell_0$ norm that counts the number of non-zero components. A proof of this lemma is provided at the end of the section.

\begin{lemma}\label{Lemma:sparsity}
Under assumptions of Theorem~\ref{Thm:Normality} and Theorem~\ref{Thm:alpha}, for any solution $\gamma$ of equation~\eqref{Eqn:M_est}, it holds with probability at least $1-p^{-c}$ for some $c>0$ that $\|\gamma\|_0 \leq C'\,s$ for some sufficiently large constant $C'$ independent of $(n,\,p,\,s)$. 
\end{lemma}

\noindent Given this lemma, our proof proceeds as follows.
Suppose there are two solutions $(\theta_1,\,\gamma_1)$ and $(\theta_2,\,\gamma_2)$ of equations~\eqref{Eqn:Za}-\eqref{Eqn:Zb}, then $\Delta\theta:\,=\theta_1-\theta_2$ and $\Delta\gamma :\,=\gamma_1-\gamma_2$ must satisfy
\begin{align*}
&\Xtil^T(X\Delta\theta + Z\Delta\gamma)=0,\quad\mbox{and}\\
&Z^T(X\Delta\theta+Z\Delta\gamma) = \lambda\, \kappa_1-\lambda\,\kappa_2,
\end{align*}
where $\kappa_1\in\partial\|\gamma_1\|_1$ and $\kappa_2\in\partial\|\gamma_2\|_1$. By solving $\Delta\theta$ from the first equation and plugging into the second, we obtain
\begin{align}
&\Delta\theta = (\Xtil^TX)^{-1}\Xtil^TZ\Delta\gamma,\quad\mbox{and}\label{Eqn:DeltaTheta}\\
&Z^T\big[I-X(\Xtil^TX)^{-1}\Xtil^T\big]Z\Delta\gamma = \lambda\,(\kappa_1-\kappa_2).\notag
\end{align}
By the definition of sub-gradients, we have
\begin{align*}
&\|\gamma_1\|_1 \geq \|\gamma_2\|_1 + \langle \kappa_2,\,\gamma_1-\gamma_2\rangle, \mbox{\quad and}\\
&\|\gamma_2\|_1 \geq \|\gamma_1\|_1 + \langle \kappa_1,\,\gamma_2-\gamma_1\rangle,
\end{align*}
implying $\langle \kappa_2-\kappa_1,\,\Delta\gamma\rangle \geq 0$ by adding them together. Putting pieces together, we obtain
\begin{align*}
\frac{1}{n}\,\|Z\Delta\gamma\|^2 \leq \frac{1}{n}\,\Delta\gamma^TZ^TX(\Xtil^TX)^{-1}\Xtil^TZ\Delta\gamma.
\end{align*}
By H\"{o}lder's inequality, we can bound its right hand side by
\begin{align*}
\frac{1}{n}\,\big|\Delta\gamma^T Z^TX(\Xtil^TX)^{-1}\Xtil^TZ\Delta\gamma\big| &\leq \frac{1}{n}\,\|\Delta\gamma\|_1 \, \|Z^TX(\Xtil^TX)^{-1}\Xtil^TZ\Delta\gamma\|_\infty\\
&\leq \|\Delta\gamma\|_1 \, \|n^{-1}Z^TX\|_{\infty,\infty}\, \|(n^{-1}\Xtil^TX)^{-1}\|_{\infty,\infty}\,\|n^{-1}\Xtil^TZ\|_\infty\, \|\Delta\gamma\|_1\\
&\leq 2\,C^2C'\,\tau\, s \|\Delta\gamma\|^2,
\end{align*}
where in the last step we have used the conditions on varies norms on the relevant matrices and $\|\Delta\gamma\|_1\leq \sqrt{2C's}\,\|\Delta\gamma\|$ since according to Lemma~\ref{Lemma:sparsity}, $\Delta\gamma$ is at most $2C's$ sparse.

Now, by combining the last two displays and the SEC (since $\Delta\gamma$ is $2C'\,s$ sparse), we obtain
\begin{align*}
\mu\,\|\Delta\gamma\|^2\leq 2\,C^2C'\,\tau\, s \|\Delta\gamma\|^2,
\end{align*}
implying $\Delta\gamma=0$ since $\mu \geq 2\,C^2C'\,\tau\, s$. Consequently, we must have $\gamma_1=\gamma_2$, and $\theta_1=\theta_2$ by applying equation~\eqref{Eqn:DeltaTheta}. Therefore, the solution of equations~\eqref{Eqn:Za}-\eqref{Eqn:Zb} is unique, which also implies the uniqueness of the solution of equations~\eqref{Eqn:Ma}-\eqref{Eqn:Mb}.

\subsection{Proof of Theorem~\ref{Thm:Normality}}
By plugging the true data generating model $Y=X\thetastar+Z\gammastar+w$ into the first constraint of problem~\eqref{Eqn:M_est_const}, we obtain
\begin{align}\label{Eqn:key_identity}
\frac{1}{n}\, \Xtil^T\Xtil( \thetahat - \thetastar) = &\, \frac{1}{n}\Xtil^T w - \frac{1}{n}\Xtil^T Z(\gammahat - \gammastar)-\frac{1}{n}\Xtil^T Z \alpha(\thetahat-\thetastar),
\end{align}
where recall that $\Xtil = X-Z\alpha$ denotes the $n\times d$ residual matrix.
By multiplying both with $(\Xtil^T\Xtil/n)^{-1}$ and using the fact that for any matrix $A\in\mb R^{m\times n}$ and vector $b\in\mb R^{n}$, 
\begin{align}\label{Eqn:Mat_Vec_Ineq}
\|Ab\|_\infty \leq \|A\|_\infty \|b\|_1 \quad\mbox{and}\quad \|Ab\|_\infty \leq \|A\|_{\infty,\infty}\, \|b\|_\infty,
\end{align}
we obtain that
\begin{align*}
\|\thetahat - \thetastar\|_\infty \leq C\,\Big(\frac{1}{n}\|\Xtil^Tw\|_1 + \tau \|\gammahat-\gammastar\|_1 +\nu\,d\,\|\thetahat-\thetastar\|_\infty \Big),
\end{align*}
where we have used the conditions that $\|(\Xtil^T\Xtil/n)^{-1}\|_{\infty} \leq C$, $\|n^{-1}\Xtil^TZ\|_\infty \leq \tau$ and $\|n^{-1}\Xtil^TZ\alpha\|_{\infty,\infty}\leq d\,\|n^{-1}\Xtil^TZ\alpha\|_{\infty}  \leq d\,\nu$. By rearranging the above inequality, we obtain
\begin{align*}
\|\thetahat - \thetastar\|_\infty \leq (1-Cd\,\nu)^{-1}\Big(\frac{1}{n}\|\Xtil^Tw\|_1 + \tau \|\gammahat-\gammastar\|_1\Big).
\end{align*}
Since $w\sim \m N(0,\,\sigma^2I_n)$ and $\|n^{-1}\Xtil^T\Xtil\|_2 \leq C$, we have that under some event $\m A$ satisfying $\mb P(\m A)\geq 1-n^{-c}$ and $c>0$, $n^{-1}\|\Xtil^Tw\|_1\leq C\,\sigma\,n^{-1}$. Consequently, under this event $\m A$, we have
\begin{align}\label{Eqn:theta_bound}
\|\thetahat - \thetastar\|_\infty \leq (1-Cd\,\nu)^{-1}\Big(\frac{C\,\sigma}{\sqrt{n}} + 2\,\rhobar\,\tau\Big),
\end{align}
where we used that fact that both $\gammahat$ and $\gammastar$ are feasible for problem~\eqref{Eqn:M_est_const} so that $\max\{\|\gammahat\|_1,\,\|\gammastar\|_1\}\leq \rhobar$. 

In the following, we will combine the bound~\eqref{Eqn:theta_bound} of $\theta$ and the optimality condition of the Lasso problem~\eqref{Eqn:Zb_const} to derive a bound for $\|\gammahat-\gamma\|_1$. By plugging this bound on $\|\gammahat-\gamma\|_1$ back into equation~\eqref{Eqn:key_identity}, we can prove the desired normal approximation for $\sqrt{n}\,(\thetahat-\thetastar)$.

To begin with, we plug in $Y=X\thetastar+Z\gammastar+w$ into problem~\eqref{Eqn:Zb_const}, and use the optimaility of $\gammahat$ and the feasibility of $\gammastar$ to obtain 
\begin{align*}
\frac{1}{2n}\,\|X(\thetahat-\thetastar) + Z(\gammahat-\gammastar) - w\|^2 + \lambda\,\|\gammahat\|_1\leq 
\frac{1}{2n}\,\|X(\thetahat-\thetastar) - w\|^2 + \lambda\,\|\gammastar\|_1.
\end{align*}
After some rearrangements, we obtain the following basic inequality,
\begin{align}\label{Eqn:BasicIneq}
\frac{1}{n}\,\|Z(\gammahat-\gammastar)\|^2 + \frac{1}{n}\, \langle \gammahat-\gammastar,\, Z^TX(\thetahat-\thetastar)\rangle \leq \frac{1}{n}\,\langle \gammahat-\gammastar,\, Z^Tw\rangle + \lambda\,\|\gammastar\|_1-\lambda\,\|\gammahat\|_1.
\end{align}
Now we bound each term separately. Using H\"{o}lder's inequality and inequality~\eqref{Eqn:Mat_Vec_Ineq}, we can bound the second term on the left hand side of this basic inequality as
\begin{align*}
\Big|   \frac{1}{n}\, \langle \gammahat-\gammastar,\, Z^TX(\thetahat-\thetastar)\rangle\Big| &\leq  \frac{1}{n}\, \|\gammahat-\gammastar\|_1\, \|Z^TX(\thetahat-\thetastar)\|_\infty\\
&\leq \frac{1}{n}\, \|\gammahat-\gammastar\|_1\, \|Z^TX\|_{\infty,\infty}\,\|\thetahat-\thetastar\|_\infty\\
&\overset{(i)}{\leq} \frac{Cd}{1- Cd\,\nu}\,\Big(\frac{C\,\sigma}{\sqrt{n}} + 2\,\rhobar\,\tau\Big)\,\|\gammahat-\gammastar\|_1\\
&\leq 2Cd\, \Big(\frac{C\,\sigma}{\sqrt{n}} + 2\,\rhobar\,\tau\Big)\,\|\gammahat-\gammastar\|_1.
\end{align*}
Here, in step (i) we used the fact $\|n^{-1}Z^TX\|_{\infty,\infty}\leq d\cdot \|n^{-1}Z^TX\|_{\infty}\leq d\cdot\max_{j=1,\ldots,d}\|n^{-1/2}X_j\|$ $\cdot \max_{j=1,\ldots,p}\|n^{-1/2}X_j\|\leq Cd$ under the column normalization condition and the bound of $\|\thetahat-\thetastar\|_\infty$ in \eqref{Eqn:theta_bound}; in the last step we used the condition that $\nu\leq (2Cd)^{-1}$.
The first term on the right hand side of basic inequality~\eqref{Eqn:BasicIneq} can be bounded as
\begin{align*}
\frac{1}{n}\,\langle \gammahat-\gammastar,\, Z^Tw\rangle \leq \frac{1}{n}\, \|\gammahat-\gammastar\|_1\, \|Z^Tw\|_\infty.
\end{align*}
Since $Z^Tw$ is a $p$-dimensional random vector, whose each element $Z_j^Tw$ has a normal distribution with standard deviation $\|Z_j\|\leq \sqrt{C}\,\sqrt{n}\,\sigma$, we obtain by a union bound argument that under some event $\m B$ satisfying $\mb P(B)\geq 1-p^{-c}$ and $c>0$, 
\begin{align*}
\frac{1}{n}\,\|Z^Tw\|_\infty \leq \sigma\,\sqrt{\frac{2C\,\log p}{n}}.
\end{align*}
Combining the last three displays, we obtain that for $\displaystyle \lambda \geq 2\sigma\,\sqrt{\frac{2C\log p}{n}} +\frac{4\sigma\,C^2d}{\sqrt{n}}+ 8Cd\,\rhobar\,\tau$, the nuisance parameter estimator $\gammahat$ satisfies
\begin{align*}
0\leq \frac{1}{n}\,\|Z(\gammahat-\gammastar)\|^2  \leq  \frac{1}{2}\,\lambda\, \|\gammahat-\gammastar\|_1 + \lambda\,\|\gammastar\|_1-\lambda\,\|\gammahat\|_1.
\end{align*}
We write $\Delta = \gammahat-\gammastar$ and decompose $\gammahat$ into $\gammahat_S+\gammahat_{S^c}$, where recall that $S$ is the support of $\gammastar$. Under this notation, we have $\Delta_S=\gammahat_S-\gammastar$ and $\Delta_{S^c}=\gammahat_{S^c}$, and the preceding display implies
\begin{align}
0&\leq \frac{1}{n}\,\|Z\Delta\|^2 \leq \frac{\lambda}{2}\,\|\Delta_S\|_1 + \frac{\lambda}{2}\,\|\Delta_{S^c}\|_1 + \lambda\,\|\gammastar\|_1 - \lambda\,\|\gammahat_S\|_1-\|\Delta_{S^c}\|_1\notag\\
& \leq  \frac{3}{2}\,\lambda\,\|\Delta_S\|_1 - \frac{1}{2}\,\lambda\,\|\Delta_{S^c}\|_1.\label{Eqn:gammahat_bound}
\end{align}
Therefore, $\Delta$ belongs to the cone $\m C$ in the REC. Now by combining REC and the preceding display, we obtain
\begin{align*}
\mu \|\Delta\|^2 \leq \frac{3}{2}\,\lambda\,\|\Delta_S\|_1 \overset{(i)}\leq \frac{3}{2}\,\lambda\,\sqrt{s}\,\|\Delta_S\| \leq \frac{3}{2}\,\sqrt{s}\,\lambda\,\|\Delta\|,
\end{align*}
where in step (i) we applied H\"{o}lder's inequality and used the fact that the size of the index set $S$ is $s$. Consequently, we obtain
\begin{align}
\|\Delta\| &\leq \frac{3}{2}\,\frac{\sqrt{s}\,\lambda}{\mu}, \qquad\mbox{and}\notag\\
\|\Delta\|_1 &= \|\Delta_S\|_1+ \|\Delta_{S^c}\|_1 \leq 4\,\|\Delta_S\|_1 \leq 4\,\sqrt{s}\, \|\Delta\|\leq 6\,\frac{s\,\lambda}{\mu}. \label{Eqn:Delta_bound}
\end{align}
Plugging this and error bound~\eqref{Eqn:theta_bound} of $\thetahat$ back into the decomposition~\eqref{Eqn:key_identity} of $\thetahat$, we obtain
\begin{align}
&\big\|\sqrt{n}\,(\thetahat-\thetastar) - \sqrt{n}\, (\Xtil^T\Xtil)^{-1}\Xtil^T w\big\|_\infty\notag \\
\leq&\, \frac{1}{\sqrt{n}}\,\|\Xtil^TZ\|_{\infty}\,\|\Delta\|_1 + \frac{2C}{\sqrt{n}}\,\|\Xtil^TZ\alpha\|_\infty \, \Big(\frac{C\,\sigma}{\sqrt{n}} + 2\,\rhobar\,\tau\Big)\label{Eqn:theta_boundKey}\\
\leq&\, \frac{6\sqrt{n}\, \tau\,s\,\lambda}{\mu} + 2C^2\,\sigma\,\nu + 4C\,\sqrt{n}\,\rhobar\,\tau\,\nu.\notag
\end{align}
yielding the claimed result.

\subsection{Proof of Theorem~\ref{Thm:alpha}}
In this proof, the meaning of constant $C'$ may be changed from line to line to simply the presentation.
By the KKT condition of the optimization problem~\eqref{Eqn:Choose_alpha}, we have
\begin{align*}
n^{-1}Z^T(X_j-Z\alpha_j) = \lambda_j\, \kappa_j,\quad \kappa_j\in\partial\|\alpha_j\|_1.
\end{align*}
By definition, the sub-gradient satisfies $\|\kappa_j\|_\infty \leq 1$, implying
\begin{align*}
\|n^{-1}\Xtil^TZ\|_\infty = \max_j\|n^{-1}Z^T(X_j-Z\alpha_j) \|_\infty \leq \max_j\lambda_j,
\end{align*}
which is the first claimed bound.

Now we prove the second bound on $\|n^{-1}\Xtil^TZ\alpha\|_\infty$. Since the $(j,\,k)$th element of this matrix satisfies
\begin{align*}
|n^{-1}(X_j - Z\alpha_j)^TZ\alpha_k| \leq \|n^{-1}Z^T(X-Z\alpha_j) \|_\infty \,\|\alpha_k\|_1,
\end{align*}
by applying the first bound, it suffices to show that $\|\alpha_j\|_1\leq 3\|\alphastar_j\|_1$ holds with high probability for all $j=1,\ldots,d$. In fact, by the optimality of $\alpha_j$ and feasibility of $\alphastar_j$ in the optimization problem~\eqref{Eqn:Choose_alpha}, we have the following basic inequality
\begin{align*}
\frac{1}{2n}\,\|X_j-Z\alpha_j\|^2 + \lambda_j\,\|\alpha_j\|_1\leq \frac{1}{2n}\,\|X_j-Z\alphastar_j\|^2 + \lambda_j\,\|\alphastar_j\|_1.
\end{align*}
Let $v_j = X_j - Z\alphastar_j\in\mb R^n$. By Assumption D, components of $v_j$ are i.i.d.~with mean zero and sub-Gaussian tails, and by the definition of $\alphastar_j$, $v_j$ also satisfies $\mb E[Z^Tv_j] = 0$.
After simple algebra, the preceding basic inequality leads to
\begin{align}\label{Eqn:Bound_alpha}
\frac{1}{n}\,\|Z(\alpha_j-\alphastar_j)\|^2 \leq \Big\langle \frac{2}{n}\,Z^Tv_j,\,  \alpha_j-\alphastar_j\Big\rangle + \lambda_j\,\big(\|\alphastar_j\|_1 - \|\alpha_j\|_1\big).
\end{align}
By applying a union bound to $p$ sub-Gaussian variables $(Z^i)^Tv_j$, we obtain that with probability at least $1-p^{-c}$ for some $c>0$, 
\begin{align}\label{Eqn:BoundCross}
\Big\|\frac{2}{n}\,Z^Tv_j\Big\|_\infty \leq \frac{\lambda_j}{2}.
\end{align}
Combining the two preceding displays, we obtain 
\begin{align*}
0\leq \frac{\lambda_j}{2}\, \|\alpha_j-\alphastar_j\|_1 + \lambda_j\,\big(\|\alphastar_j\|_1 - \|\alpha_j\|_1\big),
\end{align*}
implying $\|\alpha_j\|_1\leq 3\|\alphastar_j\|_1$ by using the triangle inequality. Finally, by applying a union bound over $j=1,\ldots,d$, we obtain that under some event $\m A$ satisfying $\mb P(\m A)\geq 1-d\,p^{-c}$, it holds that $\max_{j=1,\ldots,d}\|\alpha_j\|_1\leq 3 \,\max_{j=1,\ldots,d}\|\alphastar_j\|_1$.

Now we prove the last part of the theorem.
Let $\Sigma^\ast =\mb E[(X-Z\alphastar)^T(X-Z\alphastar)]$ denote the inverse of $\Omega^\ast$. It suffices to prove a bound on $\|n^{-1}\Xtil^T\Xtil-\Sigma^\ast\|_\infty$, which combined with the fact that $\Sigma^\ast$ is positive definite and the inequality $\|(A+\Delta)^{-1}-A^{-1}\|\leq \|A^{-1}\|^2\|\Delta\|$ with $\|\cdot\|$ being the matrix operator norm (for any symmetric matrix $B$, $\|B\|_\infty\leq\|B\|$) yields the claimed bound. 
It suffices to show that for any $(j,\, k)$, it holds with high probability that
\begin{align*}
\Big|\frac{1}{n}\,(X_j - Z\alpha_j)^T(X_k - Z\alpha_k) - E[(X_{ij}-Z^i\alphastar_j)^T(X_{ik}-Z^i\alphastar_k)]\Big| \leq C' \, \sqrt{\frac{\log p}{n}}.
\end{align*}
In fact, we have the following decomposition for the difference,
\begin{align*}
&\Big|\frac{1}{n}\,(X_j - Z\alpha_j)^T(X_k - Z\alpha_k) - E[(X_{ij}-Z^i\alphastar_j)^T(X_{ik}-Z^i\alphastar_k)]\Big|\\
&= \Big|\frac{1}{n}\, \big[Z(\alpha_j-\alphastar_j) - v_j\big]^T\big[Z(\alpha_k-\alphastar_k) - v_k\big]-E[v_j^Tv_k]\Big|\\
& \leq \Big\|\frac{1}{\sqrt{n}}\, Z(\alpha_j-\alphastar_j)\Big\|\, \Big\|\frac{1}{\sqrt{n}}\, Z(\alpha_k-\alphastar_k)\Big\| + \Big|\langle \frac{1}{n}\,Z^Tv_j,\,  \alpha_k-\alphastar_k\Big\rangle\Big|\\
&\qquad\qquad\qquad +\Big|\Big\langle \frac{1}{n}\,Z^Tv_k,\,  \alpha_j-\alphastar_j\Big\rangle\Big| + \Big|\frac{1}{n}\,\langle v_j,\,v_k\rangle-E[v_j^Tv_k]\Big|.
\end{align*}
The last term can be bounded by $C'/n$ under some event $\m B_{jk}$ satisfying $\mb P(\m B_{jk})\geq 1-n^{-c}$.
Applying bound~\eqref{Eqn:Bound_alpha} and~\eqref{Eqn:BoundCross}, we obtain that with probability at least $\mb P\big(\m A\cap \bigcup_{j\leq k}\m B_{jk}\big)\geq 1-d\,p^{-c}-d^2\,n^{-c}$, the above can be bounded by
\begin{align*}
\frac{3}{2}\,\sqrt{\lambda_j\,\lambda_k\,\|\alphastar_j\|_1\,\|\alphastar_k\|_1} + \frac{\lambda_j}{2}\, \|\alphastar_j\|_1 +\frac{\lambda_k}{2}\, \|\alphastar_k\|_1 + \frac{C'}{\sqrt{n}}\leq C'\, \sqrt{\frac{\log p}{n}},
\end{align*}
for any $(j,\,k)\in\{1,\ldots,d\}^2$, implying the claimed result.

\subsection{Proof of Theorem~\ref{Thm:convergence}}
Similar to the derivation for the error bound~\eqref{Eqn:theta_boundKey} for $\thetahat$, it can be shown that under some event $\m A$ with $\mb P(\m A)\geq 1-n^{-c}$, for any $t\geq 1$, the deviation $\|\theta^t-\thetastar - n^{-1}W\|_\infty$ satisfies (by replacing all $\gammahat$ with $\gamma^{t-1}$ and $\lambda$ with $\lambda^{t}$)
\begin{align}
\|\theta^t-\thetastar - n^{-1/2}\,W\|_\infty &\leq \tau\,\|\gamma^{t-1}-\gammastar\|_1 +
2\,C\, \nu\, \Big(\frac{C\,\sigma}{\sqrt{n}} + 2\,\tau\,\|\gamma^{t-1}-\gammastar\|_1\Big)\notag\\
&\leq 3\, \tau\,\|\gamma^{t-1}-\gammastar\|_1 + \frac{2\,C^2\, \nu\,\sigma}{\sqrt{n}}.\label{Eqn:theta_iterate_error}
\end{align}
According to a similar analysis for the error bound~\eqref{Eqn:Delta_bound} for $\Delta=\gamma^{t}-\gammastar$, it can be proved that for any $\displaystyle \lambda_t = D\,\Big\{2\sigma\,\sqrt{\frac{2C\log p}{n}} +\frac{4\sigma\,C^2d}{\sqrt{n}}+ 8Cd\,\tau\, \|\gamma^{t-1}-\gammastar\|_1\Big\}$ for $D\geq 1$, under some event $\m B$ with $\mb P(\m B)\geq 1-p^{-c}$, it holds for all $t\geq 1$ that the difference $\Delta_t =\gamma^t-\gammastar$ belongs to the cone $\m C$ defined in REC and
\begin{align*}
&\|\gamma^t-\gamma\|\leq \frac{3}{2}\,\frac{\sqrt{s}\,\lambda_t}{\mu}\quad\mbox{and}\quad\|\gamma^t-\gamma\|_1 \leq 6\,\frac{s\,\lambda_t}{\mu}. 
\end{align*}
By plugging the expression of $\lambda_t$ and rearranging the terms, we obtain the following iterative formula for the error of estimating $\gamma$,
\begin{align*}
\|\gamma^t-\gammastar\|_1 \leq \rho\, \|\gamma^{t-1}-\gammastar\|_1 + v_n,\quad t=1,2,\ldots,
\end{align*}
where $\rho=48\, CD\, s\,\tau\,\mu^{-1}<1$ and $v_n=12\, C\,D\,\mu^{-1}\,\sigma\,s\,\sqrt{2C\log p/n} + 24\, C^2D\,d\,\sigma/\sqrt{n}$.
Consequently, by solving this recursive formula we obtain that for any $t\geq 1$
\begin{align*}
\|\gamma^t-\gamma\|_1\leq \rho^t\, \|\gamma^{0}-\gammastar\|_1 + \frac{v_n}{1-\rho}.
\end{align*}
By plugging this back into the error bound~\eqref{Eqn:theta_iterate_error} for $\theta^t$, we obtain that for any $t\geq1$,
\begin{align*}
&\|\sqrt{n}\,(\theta^t-\thetastar) - W\|_\infty \leq 3\sqrt{n}\,\tau\,\rho^{t-1}\,\|\gamma^0-\gammastar\|_1 + \vep_n,\\
\mbox{with\quad}&\quad \vep_n = 36\,CD\,\frac{\sigma\,\tau\,s}{(1-\rho)\,\mu} \,\sqrt{2C\,\log p}+ 72\,C^2D\,\frac{\sigma\,\tau\,d}{1-\rho}+2\,C^2\, \sigma\,\nu.
\end{align*}

\subsection{Proof of Proposition~\ref{Prop:Relation}}
By the definitions of $\widehat{b}_1$ and $\theta^1$, we can write their difference as
\begin{align*}
\widehat{b}_1 - \theta^1_1 = \big(\widehat{\tau}_1^{-2} - \widetilde{\tau}_1^{-2}\big) \,\Big(\frac{1}{n}(X-Z\alpha_1)^T\,w  - \frac{1}{n}(X-Z\alpha_1)^T\big(X(\theta^0-\thetastar) +Z(\gamma^0-\gammastar)\big)\Big).
\end{align*}
Since $(\theta^0,\,\gamma_0)$ is the solution to the unadjusted Lasso, according to the classical results on the prediction risks for Lasso (see, for example, \cite{meinshausen2009lasso}), we can bound the second term as
\begin{align*}
\frac{1}{\sqrt{n}}\,\|X(\theta^0-\thetastar)+Z(\gamma^0-\gammastar)\|&=O_P\Big(\frac{\sqrt{s\,\log p}}{n}\Big).
\end{align*}
By applying a union bound for the maximum of Gaussian random variances, we can bound the first term as
\begin{align*}
\frac{1}{n}\big|(X-Z\alpha_1)^T\,w\big| = O_P\Big(\frac{\sqrt{\log p}}{n}\Big).
\end{align*}
According to the proof of Theorem~\ref{Thm:alpha}, we have (recall that $X\in\mb R^{n\times d}$ with $d=1$)
\begin{align*}
\Big|\frac{1}{n}\,\|X-Z\alpha_1\|^2 - \mb E\big[\|X^i-Z^i\alpha_1\|^2]\Big| = O_P\Big(\sqrt{\frac{\log p}{n}}\Big).
\end{align*}
Recall that $\widehat{\tau}_1^2 =  n^{-1}\,\|X-Z\alpha_1\|^2 + 2\lambda_1\,\|\alpha_1\|_1$. According to the proof of Theorem~\ref{Thm:alpha}, $\|\alpha_1\|_1\leq 3\|\alphastar_1\|_1$ holds with high probability and
$\lambda_j$ is of order $\sqrt{n^{-1}\log p}$. Consequently, by putting pieces together, we obtain
\begin{align*}
\big|\widehat{\tau}_1^2 - \mb E\big[\|X^i-Z^i\alpha_1\|^2]\big| = O_P\Big(\sqrt{\frac{\log p}{n}}\Big).
\end{align*}
Similarly, we can decompose $\widetilde{\tau}_1^2=n^{-1}\|X-Z\alpha_1\|^2 +n^{-1}(X-Z\alpha_1)^TZ\alpha_1$. According to the proof of Theorem~\ref{Thm:alpha}, the second term $n^{-1}(X-Z\alpha_1)^TZ\alpha_1$ is $O_P(\sqrt{n^{-1}\log p})$, implying
\begin{align*}
\big|\widetilde{\tau}_1^2 - \mb E\big[\|X^i-Z^i\alpha_1\|^2]\big| = O_P\Big(\sqrt{\frac{\log p}{n}}\Big).
\end{align*}

\noindent Combining all the pieces together, we obtain
\begin{align*}
\big|\widehat{b}_1 - \theta^1_1\big| = O_P\Big(\frac{\sqrt{s} \,\log p}{n}\Big).
\end{align*}
Since Theorem~\ref{Thm:convergence} implies $\mb E\big[|\theta^1_1 - \thetastar_1|^2\big]\sim n^{-1/2}$, yielding $O_P(\sqrt{s} \,\log p/n)=o_P(|\theta^1_1 - \thetastar_1|)$ as $\sqrt{s}\,\log p/n\to 0$.

\subsection{Proof of Lemma~\ref{Lemma:sparsity}}
By solving $\theta$ from equation~\eqref{Eqn:Za} and plugging it back into equation~\eqref{Eqn:Zb}, we obtain
\begin{align*}
\frac{1}{n}\,Z^T(I-\widetilde{P})(Y-Z\gamma) = \lambda\,\kappa,\quad\kappa\in\partial \|\gamma\|_1,
\end{align*}
where $\widetilde{P} = X(\Xtil^TX)^{-1}\Xtil^T$ is an idempotent matrix satisfying $\widetilde{P} X=X$. We can further obtain by plugging $Y=X\thetastar+Z\gamma+w$ into the above and rearranging terms that
\begin{align*}
-\frac{1}{n}\,Z^T(I-\widetilde{P})Z(\gamma-\gammastar) = \lambda\,\kappa -\frac{1}{n}\,Z^T(I-\widetilde{P}) w.
\end{align*}
Similar to the proof of Theorem~\ref{Thm:Normality}, the last term can be bounded as 
\begin{align*}
\Big|\frac{1}{n}\,Z^T(I-\widetilde{P}) w\Big| \leq \frac{\lambda}{2},
\end{align*}
with probability at least $1-p^{-c}$ (by Theorem~\ref{Thm:alpha}, $\|\widetilde{P}\|_\infty$ is bounded with high probability).
Let $\widehat{S}$ to denote the support of $\gamma$, that is, $\widehat{S}=\{j:\, \gamma_j\neq 0\}$. Then according to the property of sub-gradient for $\|\cdot\|_1$, we must have
$|\kappa_j|=1$ for each $j\in\widehat{S}$. Combining this with the preceding two displays, we obtain the following element-wise bound,
\begin{align*}
\Big|\frac{1}{n}\,\big[Z^T(I-\widetilde{P})Z(\gamma-\gammastar)\big]_j\Big| \geq \frac{\lambda}{2}, \quad j\in\widehat{S},
\end{align*}
where recall that $a_j$ denotes the $j$th element of vector $a$. Squaring and summing the last display over $j\in\widehat{S}$, we obtain
\begin{align}\label{Eqn:S_bound}
\frac{\lambda^2}{4}\,|\widehat{S}| \leq \frac{1}{n}\,\|Z_{\widehat{S}}^TZ_{\widehat{S}}\|\cdot \frac{1}{n}\,\|(I-\widetilde{P})Z(\gamma-\gammastar)\|^2,
\end{align}
where $\|A\|$ denotes the operator norm for any matrix $A$.
By the proof of Theorem~\ref{Thm:Normality}, any solution $\gamma$ satisfies $n^{-1}\, \|(I-\widetilde{P})Z(\gamma-\gammastar)\|^2 \leq n^{-1}\, \|I-\widetilde{P}\|^2\,\|Z(\gamma-\gammastar)\|^2\leq C''\,s\,\lambda^2$ for some constant $C''>0$. Since $Z$ is in a general position, we must have $|\widehat{S}|\leq n$. In addition, under the random design assumption and Assumption D, $n^{-1} \|Z_{\widehat{S}}^TZ_{\widehat{S}}\| \leq c_1\, \log p$ holds with probability at least $1-p^{-c}$ for some constant $c_1\geq 0$. Therefore, inequality~\eqref{Eqn:S_bound} implies $|\widehat{S}| \leq \widetilde{C} \, s\,\log p$ for some $\widetilde{C}>0$. This leads to an improved bound $\frac{1}{n}\,\|Z_{\widehat{S}}^TZ_{\widehat{S}}\| \leq C''$ by matrix concentration inequalities \cite{vershynin2010introduction}, which in turn implies by using inequality~\eqref{Eqn:S_bound} that $|\widehat{S}|\leq C_1 s$ for some constant $C_1>0$.

\section{Discussion}\label{Sec:Discussion}
In this paper, we proposed the Constrained Lasso (CLasso) by incorporating a zero-bias constraint with the Lasso programming. We show that the resulting estimator attains root-$n$ consistency and has an asymptotically normal limiting distribution that facilitates statistical inference in the presence of high-dimensional parameters. We also propose a globally convergent algorithm for numerically computing the CLasso estimator. Our theory indicates that the state-of-the-art de-sparsified type estimators are asymptotically equivalent to the first iterate in the proposed iterative algorithm for implementing CLasso when the algorithm is initialized at the Lasso estimator. Simulations show that our method gains encouraging improvement over the de-biased estimators. 

One unanswered open problem is that whether the estimating equations in~\eqref{Eqn:Z_est} actually correspond to the KKT condition of any (convex) $M$-estimation procedure when $\alpha \neq 0$.
As future directions, we would also like to extend the CLasso by replacing the Lasso with other non-convex penalized approaches such as MCP \cite{zhang2010nearly} and SCAD \cite{fan2001variable} that tend to incur smaller bias in estimating the nuisance parameter. It would be also interesting to extend the CLasso from regression to other high-dimensional problems, such as classification, network learning, time dependent data prediction and etc. 

\bibliography{LASSO}

\end{document}